\RequirePackage{snapshot}
\documentclass[10pt,journal,cspaper,compsoc]{IEEEtran}
\newcommand{\etal}{\textit{et al.}}
\newcommand{\ie}{\textit{i.e.}}

\pdfoutput=1

\ifCLASSOPTIONcompsoc
\else
\fi
\ifCLASSINFOpdf
   \usepackage[pdftex]{graphicx}
\else
   \usepackage[dvips]{graphicx}
\fi
\usepackage{amsmath}
\usepackage{amssymb}
\usepackage{mathptmx}
\usepackage{hhline}
\usepackage{color}
\usepackage{soul}
\usepackage{sidecap}
\usepackage{algorithm}
\usepackage{subfig}
\usepackage[noend]{algorithmic}
\def\Mpch{h^{-1} {\rm Mpc}}

\newenvironment{varalgorithm}[1]
  {\algorithm\renewcommand{\thealgorithm}{#1}}
  {\endalgorithm}
  
  \pdfoutput=1

\begin{document}
%
\title{ Felix: A Topology based Framework for Visual Exploration of Cosmic Filaments}
%
%
%
%

\author{Nithin~Shivashankar, ~\IEEEmembership{Student~Member,~IEEE}, 
Pratyush~Pranav, 
Vijay~Natarajan,~\IEEEmembership{Member,~IEEE}, 
~Rien~van~de~Weygaert,
E~G~Patrick~Bos, and Steven~Rieder.
\IEEEcompsocitemizethanks{\IEEEcompsocthanksitem N. Shivashankar is with the Department
of Computer Science and Automation, Indian Institute of Science, Bangalore,
India, 560038.
Email:~nithin@csa.iisc.ernet.in
\IEEEcompsocthanksitem P. Pranav is with Kapteyn Astronomical Institute, 
University of Groningen, Netherlands. 
Email:~pratyuze@gmail.com
\IEEEcompsocthanksitem V. Natarajan is with the Department
of Computer Science and Automation and Supercomputer Education and Research Centre, Indian Institute of Science, Bangalore,
India, 560038.
Email:~vijayn@csa.iisc.ernet.in
\IEEEcompsocthanksitem R. van~de~Weygaert is with Kapteyn Astronomical Institute, 
University of Groningen, Netherlands. 
Email:~weygaert@astro.rug.nl
\IEEEcompsocthanksitem E~G~P~Bos is with Kapteyn Astronomical Institute, 
University of Groningen, Netherlands. 
Email:~pbos@astro.rug.nl
\IEEEcompsocthanksitem S. Rieder is with RIKEN Advanced Institute for Computational Science, Japan. 
Email:~steven.rieder@riken.jp}
\thanks{}}

%
%

\markboth{IEEE Transactions on Visualization and Computer Graphics,~Vol.~X, No.~X, January~XXXX}%
{Shivashankar \MakeLowercase{\textit{et al.}}, Felix: A Topology based Framework for Visual Exploration of Cosmic Filaments}
%


\IEEEcompsoctitleabstractindextext{%
\begin{abstract}
The large-scale structure of the universe is comprised of virialized blob-like clusters, linear filaments, sheet-like walls and huge near empty three-dimensional voids. 
Characterizing the large scale universe is essential to our understanding of the formation and evolution of galaxies. 
The density range of clusters, walls and voids are relatively well separated, when compared to filaments, which span a relatively larger range. The large scale filamentary network thus forms an intricate part of the cosmic web. 

In this paper, we describe Felix, a topology based framework for visual exploration of filaments in the cosmic web.
The filamentary structure is represented by the ascending manifold geometry of the 2-saddles in the Morse-Smale complex of the density field.
We generate a hierarchy of Morse-Smale complexes and query for filaments based on the density ranges at the end points of the filaments. 
The query is processed efficiently over the entire hierarchical Morse-Smale complex, allowing for interactive visualization. 

We apply Felix to computer simulations based on the heuristic Voronoi kinematic model and the standard $\Lambda$CDM cosmology, and demonstrate its usefulness through two case studies. 
First, we extract cosmic filaments within and across cluster like regions in Voronoi kinematic simulation datasets. 
We demonstrate that we produce similar results to existing structure finders. 
Second, we extract different classes of filaments based on their density characteristics from the $\Lambda$CDM simulation datasets. 
Filaments that form the spine of the cosmic web, which exist in high density regions in the current epoch, are isolated using Felix.
Also, filaments present in void-like regions are isolated and visualized.
These filamentary structures are often over shadowed by higher density range filaments and are not easily characterizable and extractable using other filament extraction methodologies. 

\end{abstract}

}

\maketitle

\IEEEdisplaynotcompsoctitleabstractindextext

%
\IEEEpeerreviewmaketitle


\section{Introduction}
\label{sec:introduction}
\IEEEPARstart{A}{t} scales from a megaparsec to a few hundred megaparsecs\footnote{A parsec is the standard unit of measurement of distances in the cosmos. A parsec is $3.26$ times the \emph{light-year}, the distance light covers in a year. A megaparsec is a million parsecs, the typical scale of measurement of size of the large scale structures in the universe.}, the universe has a web-like appearance. 
In the \emph{cosmic web} \cite{BKP96,weybond2008}, galaxies, intergalactic gas, and dark matter have aggregated in an intricate wispy spatial pattern marked by dense compact \emph{clusters}, elongated \emph{filaments} and sheetlike \emph{walls}, and large near-empty void regions. 
The filaments, stretching out as giant tentacles from the dense cluster nodes, serve as transport channels along which mass flows towards the clusters. 
They surround the flattened walls, which are tenuous, membrane-like features in the cosmic mass distribution. 

All structures and objects in the universe emerged out of primordial fluctuations that were generated during the \emph{inflationary era}, 
moments after its birth, as the universe underwent a rapid phase of expansion \cite{Guth1981,Linde1982a,Linde1982b}. The quantum fluctuations generated during this phase manifest themselves as fluctuations in the temperature of the cosmic microwave background \cite{Spergel2007,Komatsu2011,ade2013planck}. 
The gravitational growth of these density and velocity perturbations has resulted in the wealth of structure that we see in the Universe. 
The web-like patterns mark the transition phase from the primordial Gaussian random field to highly nonlinear structures that have fully collapsed into halos and galaxies. 
As our insight into the complex structural pattern of the cosmic web has increased rapidly over the past years, it has become clear that the cosmic-web contains a wealth of information on a range of cosmological and astronomical aspects and processes. 

An important illustration of the cosmological significance of the cosmic web concerns its dependence on the nature of 
dark energy and matter, the dominant but as yet unidentified forms of energy and matter in the Universe. 
One telling example of this is the recent realization that cosmic voids are sensitive and useful probes of the nature of dark energy and dark matter and testing grounds for modified gravity theories \cite{ryden96,ryden01,PL07,LW2010,BWDP12,LW12,clampitt13,SPWW14}. 
As the cosmic web is first and foremost defined and shaped by the gravitationally dominant dark matter, it would be of considerable importance to be able to obtain detailed maps of dark matter distribution. 
In recent years, great strides have been made towards this goal as gravitational lensing of distant galaxies and objects by the dark matter have enabled an increasingly accurate view of its spatial distribution \cite{Tyson2000,KneibNatarajan11}. 
Initial efforts
concentrated on the detection and mapping of the deep potential wells of the nodes in the cosmic web, \ie{}, of galaxy clusters.
Recent results have opened the path towards the mapping of filaments via their lensing effect on background 
sources \cite{DWCF12}. The identification of the structural components of the cosmic web is also important for our understanding 
of the relation between the formation, evolution, and properties of galaxies and the structural environment of the cosmic web. 
A direct manifestation of this is the generation of the angular momentum of galaxies. This is a product of the torqueing by the large-scale tidal force field \cite{hoyle51,peebles69,white84}. 
While these force fields are also the agents for the formation and shaping of 
filaments, we would expect that this results in the alignment of the spin axis of galaxies with respect to cosmic filaments 
\cite{leepen2000,JWA10,TSS13}. 

The identification, description, and characterization of the elements of the cosmic web is a non-trivial problem. Several 
characteristics of the mass distribution in the cosmic web have made it an extremely challenging task to devise an 
appropriate recipe for identifying them: 

\textbf{a)} The cosmic web is a complex spatial pattern 
of connected structures displaying a rich geometry with multiple morphologies and shapes. 

\textbf{b)} There are no well-defined structural objects at a single spatial scale or within a specific density range. 
Instead, elements of cosmic web are found at a wide range of densities and spatial scales. This is a consequence of 
the hierarchical evolution of structure formation in the universe, such that smaller high-density 
structures merge to form larger objects. 

\textbf{c)} There is a clear anisotropy in the structures of the cosmic web, 
a consequence of gravitational instability. The structures in the 
cosmic web exhibit elongated and flattened characteristics.

The attempts to analyze the structure of the cosmic web have a long history. The absence of an 
objective and quantitatively accurate procedure for identifying and isolating the components of the 
cosmic web has been a major obstacle in describing it. In recent years, more elaborate and advanced techniques 
have been developed to analyze and describe the structural patterns in the cosmic web. Nonetheless, a consensus on 
the proper definition of filaments is yet to be achieved. In the subsequent subsection~\ref{sec:relatedwork} we present a short account of the available techniques and the definitions on which they are based. 


\subsection{Related Work}
\label{sec:relatedwork}
Statistical measures such as the auto-correlation function \cite{Pee80} of the matter distribution 
in the web have been the mainstay of cosmological studies over many decades. However, while this 
second-order measure of clustering does not contain any phase information (one may e.g. always 
reproduce a distribution with the same 2nd order moments and random Fourier phases), the auto-correlation 
function is not sensitive to the existence of complex spatial patterns. Higher order correlation 
functions only contain a very limited amount of such structural information, while in practical 
observational circumstances, it quickly becomes cumbersome to measure them as the magnitude of the 
error increases drastically with increasing order. 

The first attempts towards characterizing complex geometric patterns in the galaxy distribution 
mainly involved heuristic measures. Early examples of techniques addressing the global connectivity of structure 
in the Universe are percolation analysis \cite{ShaZel83} and the minimum spanning tree of the spatial galaxy 
distribution \cite{BBS85,alpaslan14}. While these are useful global descriptions, they do not capture and describe local characteristics of the mass distribution. 


More elaborate and advanced techniques have been developed in recent years. Several of these methods 
apply sophisticated mathematical and visualization techniques, involving geometric and topological 
properties of the cosmic mass distribution. There are a multitude of different methods for detecting 
filaments, based on a range of different techniques. We may recognize several categories of techniques. 

One class of methods seeks to describe the local geometry on the basis of the Hessian of the density field 
\cite{aragon07a,aragon07b,bond10,CWJ13} or closely related quantities such as the tidal force field 
\cite{HPCD07,forero09} or the velocity shear field \cite{libeskind12,CWJ13}. The Hessian provides 
direct information on the local shape and dynamical impact of the corresponding field. The morphological elements 
of the cosmic web are identified by connecting the areas within which a specific range of anisotropies is registered. 

These studies concentrate on a single scale by appropriately smoothing the field, and 
do not consider the multi-scale nature of the cosmic mass distribution. The Hessian based Nexus/MMF technique \cite{aragon07b}, which was perfected into a versatile and parameter-free method \cite{CWJ13}, 
implicitly takes into account the multi-scale nature of the web-related fields. It accomplishes this by a 
scale-space analysis of the fields. At each location the optimal morphological signal is extracted via the 
application of a sophisticated filter bank applied to the Hessian of the corresponding fields in scale space. 
The application of this machinery has enabled thorough studies of the hierarchical evolution and buildup of 
the cosmic web \cite{cautun2014}. 

A promising and highly interesting recent development has opened up the path towards dynamical 
analysis of the evolving mass distribution in full six-dimensional phase-space (in which the 
position of each mass element is specified by its space coordinates and velocity/momentum). In the 6D phase space, 
the cosmic mass distributions defines a 3D sheet. Independently, three groups arrived at tessellation 
based formalisms that exploit the evolving structure and folding of the \emph{phase space sheet} in phase space 
\cite{shandarin11,abel11,neyrinck12} (also see e.g. \cite{shandarin12}). The number of folds of the phase space sheet at 
a given location indicates the number of local velocity streams, and forms a direct indication of the morphology of the local 
structure. Interestingly, the resulting characterization of the web-like distribution, the Origami formalism 
of Neyrinck \cite{neyrinck12} for example, appears to resemble that of the Nexus/MMF formalism \cite{cautun2014}.

An entirely different class of techniques, based on statistical methods, have also been used to recover the filamentary patterns in the Universe. The key idea behind these techniques is to treat the galaxy distribution as a Markov point process. Within this class, particularly worth mentioning is the Bisous or the Candy model, which has been used by Stoica \etal{} \cite{stoica07} to detect the filamentary network \footnote{See Stoica \etal \cite{stoica05}, for a detailed description of this object point process which is used to characterize an observed point distribution by a fitting procedure using global optimization techniques.}. 
In this model, one places a random configuration of interacting geometric cylindrical objects on the point process to detect filamentary structure.
It has been developed into a versatile, statistically solid yet computationally challenging 
formalism for the identification of filaments in a spatial point distribution, such as N-body simulations and 
galaxy redshift surveys \cite{tempel13}. An additional example of a method involving statistical analysis of 
a geometric model is that of Genovese \etal{} \cite{GPVW10}, which seeks to describe the filamentary patterns of the cosmic web in a non-parametric way by recovering the medial axis \cite{blum1967} of the point-set of galaxies. 

The fourth major class of methods, the one which we will also pursue in this paper, exploits the 
topological structure of the cosmic mass distribution. The fundamental basis of these methods is 
Morse theory \cite{Mil63}. The geometric structure of the {\it Morse-Smale complex} \cite{EHZ01} naturally 
delineates the various morphological components on the basis of the connections between the 
critical points of the density fields and the higher dimensional cells that are incident on the 
critical points. Various Morse theory based formalisms have been applied to the identification of components of 
the cosmic web. One of the first applications concerned the detection of voids in the cosmic density field. The Watershed Void 
Finder \cite{PWJ07} identifies these with the watershed basins around the density minima. The SpineWeb procedure 
\cite{APWS10} extended the watershed transform towards the detection of the full array of structural components, filaments, walls 
and voids. These techniques use a user-defined filter to incorporate the multi-scale structure of the cosmic density 
field. 

A natural topological means to address the multi-scale topological structure emanates from 
the concept of persistence \cite{ELZ02}. It provides a natural recipe for detecting and quantifying
the components of the cosmic web in a truly hierarchical fashion. Sousbie \cite{Sousbie1,Sousbie2} has exploited and 
framed this in an elegant and impressive framework, the {\rm DisPerSE} formalism. Following the construction 
of the Morse-Smale complex, they proceed to simplify it. The simplification proceeds by 
canceling pairs of critical points iteratively, where each pair represents a structure in the cosmic web. Topological 
persistence is invoked to order the critical point pairs. However, this measure of importance is not unique, 
and one may consider alternatives, dependent on the specific interest and purpose. 

In effect, to tackle similar issues in other visualization areas, a range of variations have 
been proposed in other studies \cite{Gunther2012,Reininghaus2011,Weinkauf2009}. 
Weinkauf \etal{} \cite{Weinkauf2009} describe the concept of separatrix persistence, where 
they compute the strength of separation of points on a separatrix curve (in 2D) connected to 
a saddle as the sum of the absolute differences of function values of the saddle and the extrema connected to it. 
This concept is extended to 3D separating sheets by Gunther \etal{} \cite{Gunther2012}. 
Reininghaus \etal{} \cite{Reininghaus2011} develop the concept of scale-space persistence where they accumulate the absolute difference in function value measure of critical points across a hierarchy of derived functions. 
The set of derived functions are generated by smoothing the function using a family of Gaussian kernels of increasing variances. 
This is similar to the Multi-scale Morphology Filter Nexus/MMF \cite{aragon07b,CWJ13} described above. 
Both methods adopt the scale space formalism as the first step to detect features at multiple scales. 
However, scale space persistence and separatrix persistence, disregard specific density regimes of interest and are potentially inappropriate when small scale features with specific density characteristics are of interest. 

\subsection{Present study: contributions}

In the present study, we describe and introduce a technique for the identification 
of filaments based on the topological characteristics of the density field. 
A key aspect of the proposed technique is its interactive nature, involving a tunable 
density parameter. Specifically, we describe the following contributions: 

	\textbf{a)} We describe Felix\footnote{The name Felix is formed from 
	an abbreviation of \textit{Fil}ament \textit{ex}plorer.}: a topology based framework 
	for visual exploration of filaments in the cosmic web.
	In particular, we develop a query framework to extract filamentary
	structures from a hierarchy of Morse-Smale complexes of the density field. 
	The filaments in Felix are parameterized 
	by the density values of the maxima and the 2-saddles that define 
	them. 
	
	
	\textbf{b)} Using Felix, we develop a semi-automatic structure finder that classifies 
	galaxies as cluster/filamentary or not. 
	We demonstrate its efficiency through two tests. 
	First, using the Voronoi Kinematic model as a benchmark, we 
	demonstrate that we are able to recover the classification with high efficiency. 
	Second, we show that the classifications are quantitatively comparable to, 
    and in several cases better than, existing classifiers. 
	
	\textbf{c)}  We investigate the nature of filaments in two different 
	density regimes from the $\Lambda$CDM simulations. The first concerns filaments 
	in the high density regions around compact dense clusters, which
	are known to function as the transport channels along which matter moves into
	the clusters. A second regime concerns the tenuous low-density filaments 
	found in low-density void regions. 
	In the supplemental document, we present an additional experiment, where we investigate 
	the nature of three classes of filaments in a relatively cleaner region 
	of a $\Lambda$CDM dataset. 
	
	\textbf{d)} In the supplemental document, we describe an efficient structure based volume 
	rendering enhancement routine that allows us to highlight 
	the density distribution in regions that are close to the selected 
	features. 



The distinction between noise and significant structures is often ill-defined, and at occasions noise may be confused with genuine structures in the hierarchically evolved mass distribution (see Figure~\ref{fig:persplot_demo_figs} and the caption thereof for an illustration). 
This problem is more pronounced when one studies the properties of tenuous filaments and walls in 
low density void-like regions. 
For the understanding of the formation and 
evolution of galaxies in such regions, we need to assess the possible 
dependence of galaxy and halo properties on the morphology and density of the 
local environment. This must be based on the successful extraction of filaments 
in low density regions and the correct identification of galaxies associated 
with them. 
In view of this, we include an interactive handle on the density regimes so that one can 
concentrate on and probe structures in specific density regimes.

The remainder of the paper is organized as follows. 
Section~\ref{sec:background} introduces necessary background material. 
Section~\ref{sec:methodology} describes Felix. 
Section~\ref{sec:modeldescription} introduces the cosmological datasets used in the experiments. 
Section~\ref{sec:Experiments} discusses the application of Felix as a structure finder and as a tool for exploring filaments in different density regimes. 
Section~\ref{sec:conclusion} concludes the paper by summarizing the main results and possible future directions.

\section{Background}
\label{sec:background}
This section reviews relevant background on Morse functions, the Morse-Smale 
complex, and topological simplification. This is a necessary prerequisite for understanding the definition of filaments and extraction methods described in the subsequent sections. 

\subsection{Morse theory and the Morse-Smale complex}

Let $f:\mathbb{M} \rightarrow \mathbb{R}$ be a real-valued scalar function defined on a manifold $\mathbb{M}$. 
Critical points of $f$ are points of $f$ where the gradient of $f$ vanishes \ie{}, $\nabla f = 0 $. 
Morse theory is the study of the relationship between the topology of level sets of scalar functions and the critical points of the function. 
The function $f$ is said to be a \emph{Morse function} if all of its critical points are non-degenerate \ie{}, the \emph{Hessian} of $f$, equal to the matrix of second order partial derivatives, is non-singular.
The non-degeneracy condition imposes a locally quadratic form for $f$ within a small neighborhood of its critical points. In other words, using a coordinate transformation, the function near a critical point $p$ of the $n$-dimensional manifold $\mathbb{M}$ can be written as $f_p(x) = f(p) \pm x_1^2 \pm x_2^2 \pm ... \pm x_n^2.$
The \emph{index} of $p$ is equal to the number of negative quadratic terms in the above expression. 
In 3D, the index 0 corresponds to minima, the index 1 corresponds to 1-saddles, the index 2 corresponds to  2-saddles, and the index 3 corresponds to maxima.
An \emph{integral line} is a maximal curve in the domain, whose tangent aligns with the gradient of $f$ at every point. 
The function $f$ increases along the integral line and its limit points are the critical points of $f$. 
\begin{figure}
\centering
\includegraphics[width=0.3\textwidth]{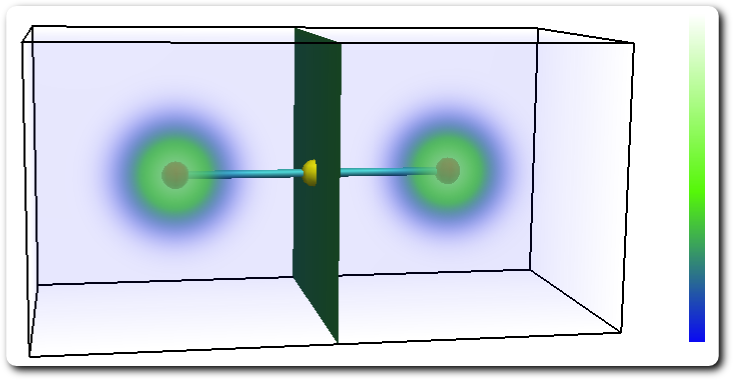}
\caption{Ascending manifolds of a 2-saddle (yellow sphere). The scalar function is a sum of two 3D Gaussians centered on either side of the volume. The two arcs incident on the 2-saddle constitute the ascending manifold and terminate at the two maxima (red spheres) of the scalar function. 
}
\label{fig:max-sad-concept}
\vspace{-15pt}
\end{figure}

The set of all integral lines that originate at the critical point $p$ together with $p$ is called the \emph{ascending manifold} of $p$.
Similarly, the set of all integral lines that terminate at the critical point $p$ together with $p$ is called the \emph{descending manifold} of $p$. 
The ascending manifolds (similarly, the descending manifolds) of all critical points partition the domain. 
The ascending manifold of a critical point with index $i$ has dimension $n-i$, where $n$ is the dimension of the domain. 
Thus, the ascending manifold of a minimum is a three dimensional cell, the ascending manifold of a 1-saddle is a two dimensional sheet, the ascending manifold of a 2-saddle is a one dimensional arc (see Figure~\ref{fig:max-sad-concept}), and the ascending manifold of a maximum is equal to the maximum. 
The converse is true for the descending manifold i.e., the descending manifold of a critical point with index $i$ has dimension $i$. 
The \emph{Morse-Smale complex} is a partition of the domain into cells formed by the collection of integral lines that share a common source and a common destination. 
The function $f$ is called a \emph{Morse-Smale function} if the ascending and descending manifolds of all pairs of critical points intersect only transversally i.e., if the ascending and descending manifolds of two critical points intersect, then the intersection has dimension exactly equal to the difference in the indices of the two critical points.
The critical points, referred to as \textit{nodes}, together with the 1-manifolds that connect them, referred to as \textit{arcs}, form the 1-skeleton of the Morse-Smale~complex, which is referred to as the \textit{combinatorial structure} of the Morse-Smale complex. 

\subsection{Morse-Smale complex simplification}
\label{subsec:topo_simp}

\begin{figure}
  \centering
  \includegraphics[width=0.5\textwidth]{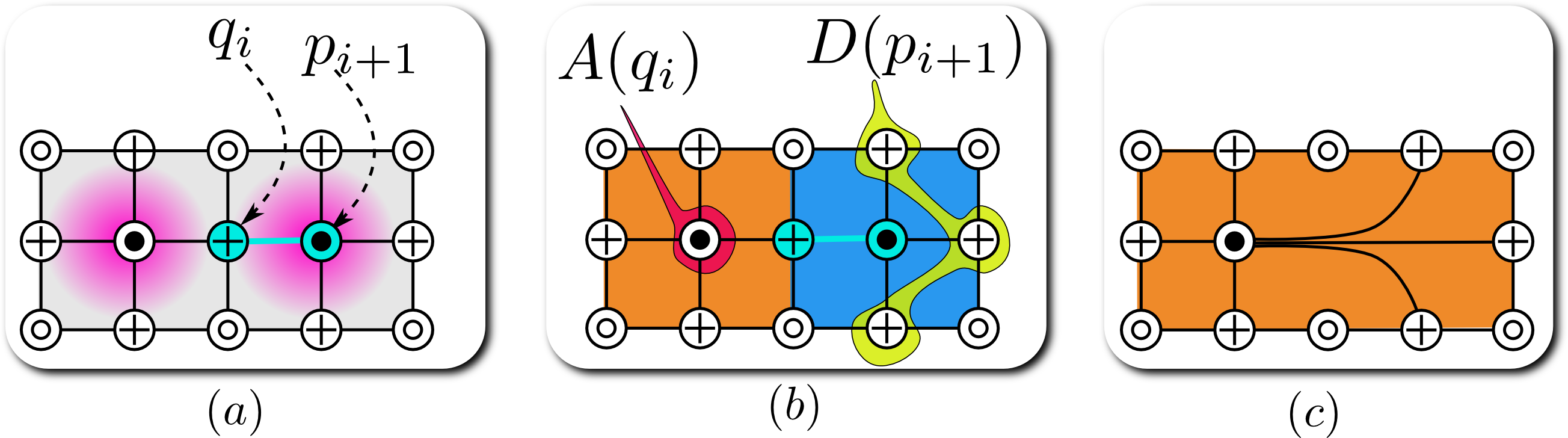}
\caption{Topological cancellation of a pair of critical points in a 2D Morse-Smale complex. 
(a)~Morse-Smale complex of the function shown in Figure~1 restricted to a 2D slice. Maxima are denoted by $\odot$, saddles by $\oplus$, and minima by $\circledcirc$. A pair ($p_{i+1}, q_i$) of critical points connected by a single arc is scheduled to be canceled. 
(b)~$D(p_{i+1})$ is the set of surviving index $i$ critical points connected to $p_{i+1}$ and $A(q_i)$ is the set of surviving index $i+1$ critical points connected to $q_i$.
(c)~Combinatorial realization: connect all critical points $D(p_{i+1})$ to those in $A(q_i)$. Geometric realization: merge the descending manifold of $p_{i+1}$ with those of critical points in $A(q_i)$. Merge the ascending manifold of $q_i$ with those of critical points in $D(p_{i+1})$.}
\label{fig:simplification}
\vspace{-15pt}
\end{figure}

The Morse-Smale complex may be simplified by repeated application of the {\it topological cancellation} procedure.
Topological cancellation eliminates a pair of critical points in the Morse-Smale complex connected by a single arc. 
The resulting complex is representative of a smoother version of $f$, which may be obtained by a local smoothing operation within an infinitesimal neighborhood of the arc.
The canceled pair of points are no longer critical in the smoother version of $f$. 
By repeatedly applying topological cancellations, one obtains a simpler Morse-Smale complex, where undesirable pairs of critical points are eliminated. 
This simplified Morse-Smale complex is representative of the function $f$ with several smoothing operations applied to it. 
Pairs of critical points that are multiply connected in the Morse-Smale complex, referred to as \emph{strangulations}, cannot be simplified by topological cancellation.

The cancellation procedure mandates changes to the combinatorial structure and ascending/descending manifolds of critical points that survive the cancellation. 
These changes are respectively referred to as the combinatorial and geometric realization of the cancellation. 
Let $p_{i+1}$ and $q_i$ denote the pair of critical points to be eliminated with $i+1$ and $i$ being the respective Morse indices.  
The combinatorial realization proceeds by first removing $p_{i+1}$ and $q_i$ as well as all arcs incident upon them.
Next, new arcs are introduced between every surviving index $i$ critical point that was connected to $p_{i+1}$ and every surviving index $i+1$ critical point that was connected to $q_i$.
The geometric realization merges the descending manifold of $p_{i+1}$ with the descending manifold of each surviving index $i+1$ critical point connected to $q_i$. 
Analogously, the ascending manifold of $q_i$ is merged with the ascending manifold of each surviving index $i$ critical point connected to $p_{i+1}$. 
Figure~\ref{fig:simplification} illustrates both realizations of a topological cancellation procedure applied to the Morse-Smale complex of a two-dimensional slice of the function shown in Figure~\ref{fig:max-sad-concept}.



The ordering of cancellation pairs plays a crucial role in determining the resulting structure of the Morse-Smale complex and its geometry. 
To simplify the Morse-Smale complex, pairs of singularly connected critical points having the least absolute difference in function value are iteratively canceled.
This approach is equivalent to the notion of topological persistence \cite{EHZ01,ELZ02} for 2D Morse-Smale complexes, but not necessarily for 3D Morse-Smale complexes \cite{guenther13}. 

\subsection{The hierarchical Morse-Smale complex}

\begin{figure}
\centering
\includegraphics[width=0.5\textwidth]{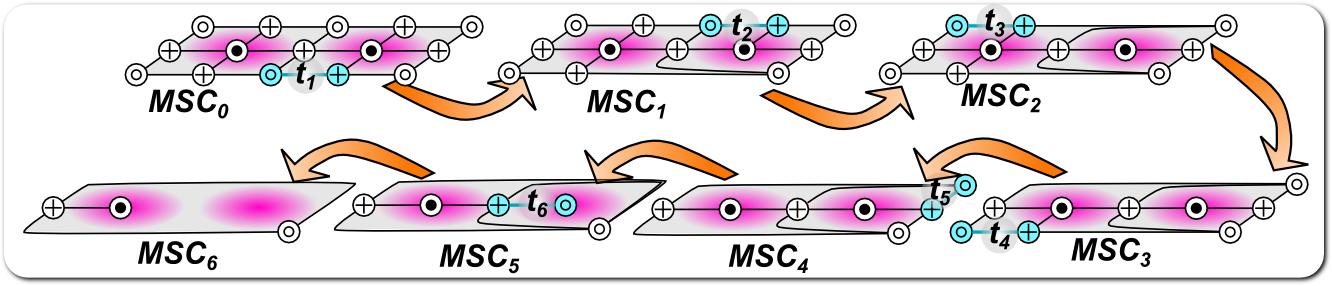}
\caption{
Hierarchical Morse-Smale complex. A family of Morse-Smale complexes generated by iteratively canceling pairs of critical points. $MSC_0$ is the Morse-Smale complex of a 2D equivalent of the function shown in Figure~\ref{fig:max-sad-concept}. It is simplified to generate a coarser version, $MSC_1$, by canceling a pair of critical points (cyan) connected by a single arc and having least absolute difference in function value $t_1$. Successive versions $MSC_i$ are computed similarly by selecting arcs so that $t_0 = 0 \leq t_1 \leq \ldots \leq t_6$.}
\label{fig:hierarchical_mscomplex}
\vspace{-15pt}
\end{figure}

A sequence of cancellations results in a hierarchical sequence of Morse-Smale complexes $MSC_0, MSC_1, ..., MSC_n$, where each Morse-Smale complex is a simpler version of the preceding Morse-Smale complex containing fewer critical points. 
Morse-Smale complex $MSC_i$ is said to be coarser than $MSC_j$ if $i>j$ and finer if $i<j$. 
The version index $i$ enumerates the Morse-Smale complexes in the hierarchy. Each non-zero version of the Morse-Smale complex, $MSC_i$, is associated with the absolute difference in function value, $t_i$, of the pair of critical points canceled in the preceding version, $MSC_{i-1}$. As each iteration selects the pair of critical points with the least absolute difference in function value, the sequence of $t_i$'s is monotonically increasing i.e. $(t_0=0 ) \leq  t_1 \leq t_2 \ldots \leq t_n$. For completeness of the sequence, $t_0$ is set to zero. 
Figure~\ref{fig:hierarchical_mscomplex} illustrates a hierarchy of Morse-Smale complexes of a 2D equivalent of the function shown in Figure~\ref{fig:max-sad-concept}.

It is not necessary to explicitly store all versions of the hierarchy of Morse-Smale complexes. 
Instead, the combinatorial representation of only $MSC_0$ is computed initially. 
Subsequently, $MSC_i$ can be obtained from a finer $MSC_{i-1}$ by performing topological cancellation (see Section~\ref{subsec:topo_simp}). 
Analogously, $MSC_i$ is obtained from a coarser $MSC_{i+1}$ by applying the inverse operation of a topological cancellation.

\section{Methodology}
\label{sec:methodology}
Exploring the filamentary patterns of the cosmic web is challenging because of the large range of the spatial scales and density range it exhibits. 
A proper characterization should also account for the hierarchical nature of structures, which adds considerable challenges to the task.
Though there exist different notions of filaments, the primary evidence relied upon for extraction and analysis is most often visual.
It is therefore not surprising that structure finding methods often visually verify results by 
superimposing the extracted structures upon visualizations of the density field or the particle distribution. 
However the visualization plays a role only after structure extraction 
process in these methods. 
We differ in this respect by providing the capability to interact with the structure finding procedure and extract 
structures that are visually relevant. To accomplish such a visual 
exploration framework, a succinct model of filament definition, an 
efficient representation of hierarchical structures, and an appropriate query mechanism that supports the extraction of these structures are paramount. 
The following exposition details our framework on these terms.


\subsection{Density estimation and filament modeling}

Cosmological simulations are N-body particle 
experiments that simulate structure formation and evolution by tracing 
positions of the particles under the influence of physical laws. 
In the observational reality, the information about structures in the cosmos 
comes through observing the galaxies. The galaxies can be treated as 
particles also for the purpose of analysis in the context of large 
scale structures.

\begin{figure}
\subfloat[]{\label{fig:2gauss_persplot}
\includegraphics[width=0.22\textwidth]{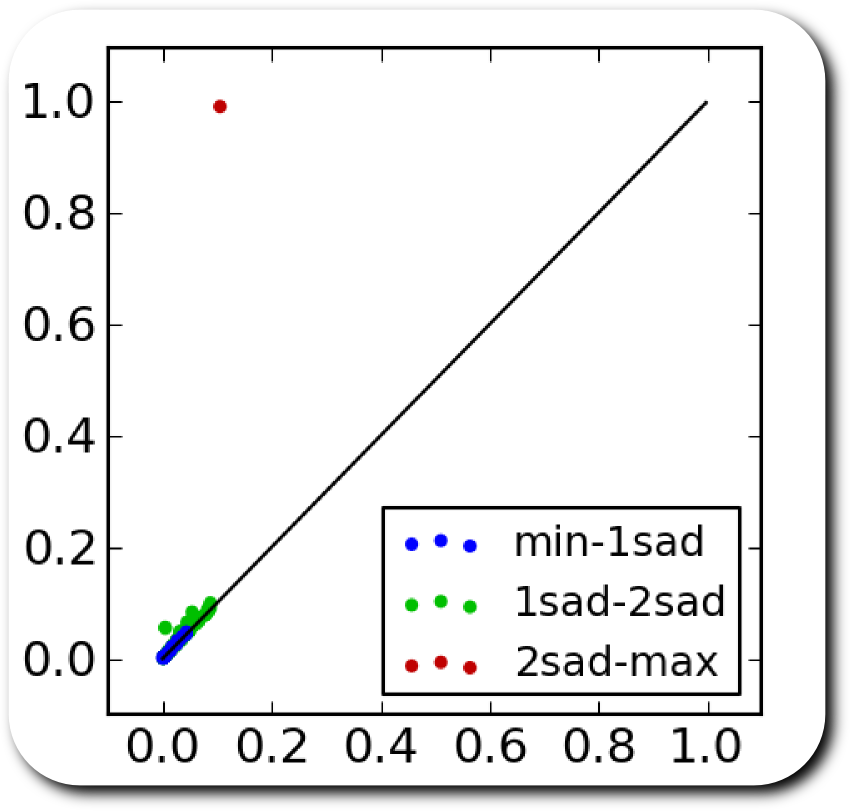}}
\subfloat[]{\label{fig:vk_new_3_persplot}
\includegraphics[width=0.22\textwidth]{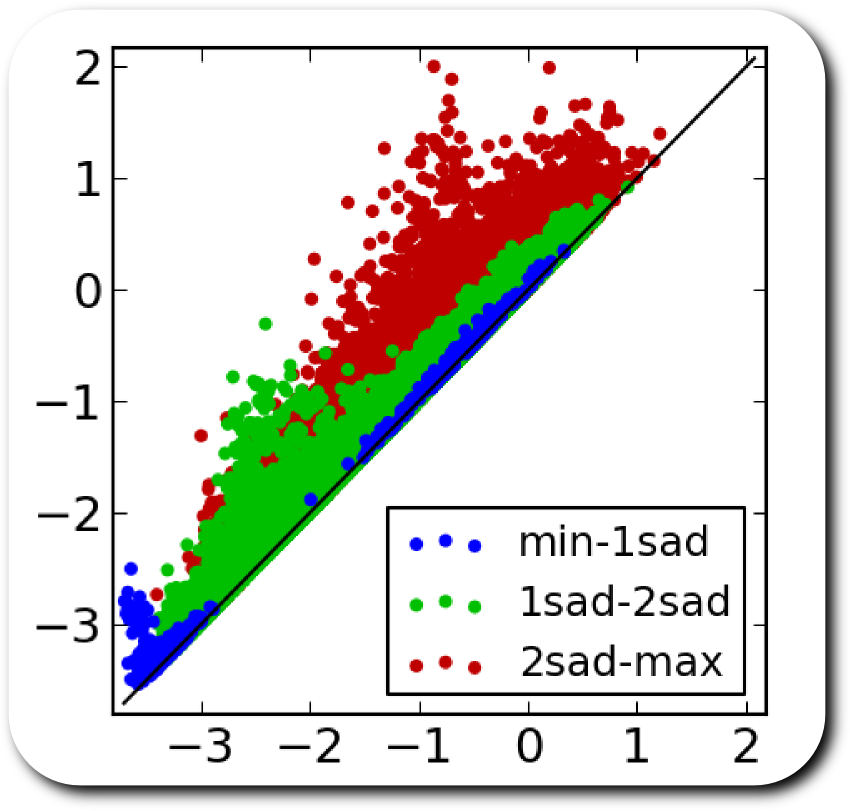}}
\caption{ \protect\subref{fig:2gauss_persplot} A scatter plot of the function values of the canceled critical point pairs for the function shown in Figure~\ref{fig:max-sad-concept}. A 2-saddle-maximum pair is the only pair that is far removed from the diagonal. This corresponds to cancellation of the 2-saddle with a maximum that represents one of the Gaussians in Figure~\ref{fig:max-sad-concept}. Other pairs close to the diagonal represent insignificant features that manifest due to the added Gaussian noise as well as sampling noise. 
\protect\subref{fig:vk_new_3_persplot} A scatter plot of the function values of the canceled critical point pairs for the Voronoi-Kinematic dataset B (see Section~\ref{subsec:voronoi-models}). No discernible separation of points is seen, though there are many points that are far removed from the diagonal. Thus, no clear global simplification threshold may be used for filament extraction. }
\label{fig:persplot_demo_figs}
\vspace{-15pt}
\end{figure}

The input to Felix is the logarithm of the density scalar field on the given 3D domain of interest. 
The domain could be 3D structured grids or tetrahedral meshes, with the density
specified on the vertices of the grid/mesh.
We find that the logarithm of the density field, instead of the density field itself, resolves the structures 
with more visual clarity. 
This has also been independently established in an earlier work \cite{CWJ13}.
Additionally, the input may be specified as a distribution of particles within a 3D region of interest. 
This could be a snapshot from a cosmological simulation, or galaxies in real observational data. 
We use the Delaunay tessellation field estimator (DTFE) \cite{DTFE,weyschaap09} to 
estimate the density of the input particles in the 3D region of interest. 
This procedure begins by computing the periodic Delaunay triangulation\cite{CGAL} on the points (simulation particles or galaxies). 
Next, the density at each vertex of the triangulation is estimated by the inverse of the volume of the tetrahedra incident upon it. Finally, the density is linearly interpolated onto the edges, faces, and tetrahedra of the Delaunay triangulation to yield a piecewise linear density function on the domain. 
The periodic Delaunay triangulation, computed by the DTFE procedure, is used to represent the domain. 
 
The Morse-Smale complex of the logarithm of the density field is computed. 
Filaments are modeled as the ascending manifolds of 2-saddles of the Morse-Smale complex. 
These arcs represent paths of steepest descent from the two maxima merging 
at the 2-saddle. This 2-saddle represents the lowest density point along the 
arcs connecting the two maxima. A schematic illustration of this is 
presented in Figure~\ref{fig:max-sad-concept}.
There are many algorithms available in the literature to compute the 3D Morse-Smale complex. The algorithms are primarily based on either the quasi Morse-Smale complex formulation~\cite{Gyulassy2007,Edelsbrunner2003} or Forman's~\cite{forman01} discrete Morse theory\cite{Gyulassy2008,VRobins2011,SSN2012,SN2012}. 
We use a parallel algorithm based on the latter approach \cite{SN2012} resulting in fast computation even for large datasets. 

The density field is rarely smooth and several local maxima obscure a 
view of the larger scale behavior of the density field. This is 
especially true if the density field is computed on the raw particle 
distribution, where the density field tends to be spiky and with a lot 
of fluctuations in the high density cluster-like regions.
The Morse-Smale complex is simplified by iteratively canceling pairs of singularly connected critical points with least absolute difference in function value to generate a hierarchy of Morse-Smale complexes. 

In most applications, a specific version of the Morse-Smale complex from the hierarchy is chosen based on a perceptibly clear separation of noise and features. 
One way to choose such a threshold separating noise and feature is by using a scatter plot of the function values of canceled critical point pairs (see Figure \ref{fig:persplot_demo_figs}) where the lower function value among the pair corresponds to the x-coordinate and the higher function value corresponds to the y-coordinate. 
In datasets where topological features are well separated (see Figure \ref{fig:2gauss_persplot}), pairs representing significant features appear far away and isolated from the diagonal. 
In such cases, the coarsest Morse-Smale complex version wherein the insignificant pairs are removed is selected for feature analysis/extraction. 
However, this strategy is not easily applicable to cosmology datasets (see Figure \ref{fig:vk_new_3_persplot}). A well defined separation is rarely discernible, though there are many scatter points that are far removed from the diagonal. 
Hence, we drop the assumption that we must work with a specific version of the Morse-Smale complex. 
Instead, we query for features across all Morse-Smale complexes in the hierarchy, as discussed in the following sub-section.



\subsection{Density range based filament selection}

Cosmic filaments exhibit a large range of variation in their density characteristics. Indeed, one expects filaments to be present both 
in void like regions and between cluster like regions. While strong dense filaments in between clusters define the spine of the 
Cosmic Web, in the hierarchically evolving mass distribution we encounter a wide spectrum of ever more tenuous filaments on smaller 
mass scales. Small filaments define the directions of mass inflow into galaxies, and form a crucial component in the formation 
of galaxies \cite{codis2012}. Even more tenuous are the systems of filaments stretching over the hollows of voids, often 
conspicuously aligned along the direction defined by neighbouring superstructures. The understanding of this network is tightly 
related to the issue of the ``missing" dwarf galaxies in voids \cite{peebles2001}. While this illustrates the complexity of 
the multiscale filigree of filaments in the Cosmic Web, we follow a strategy in which we focus our attention on 
specific aspects and details of the cosmic web. Dependent on the identity of objects and structures of interest, we 
wish to be able to zoom in on to the corresponding filamentary network. This is largely dependent on the mass scales 
of the objects involved, and the density values of the corresponding filament generating density peaks \cite{Aragon2010,cautun2014}. 

\begin{figure}
\centering
\includegraphics[width=0.4\textwidth]{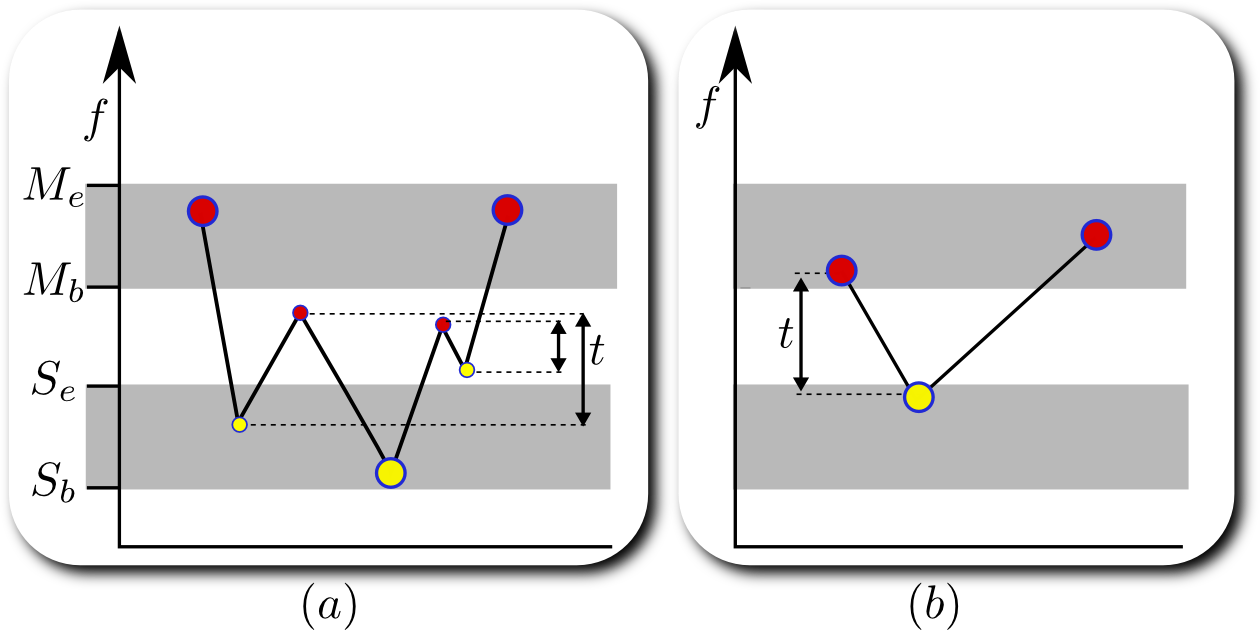}
\caption{ (a) Filaments are modeled as the ascending paths of 2-saddles connecting two extrema. The 2-saddles are filtered based on the range constraints $[M_b,M_e]$ and $[S_b,S_e]$ on the highest and lowest values respectively along the ascending paths. The highest values along the 2-saddle's ascending manifold are at extrema and the lowest value is at the 2-saddle. 
The function along the paths needs to be simplified as it is rarely smooth. In the illustration, a simplification threshold of $t$ reveals a filament with appropriate density characteristics. 
However, imposing such a threshold uniformly will cause another filament (b) having the required density characteristics to be destroyed. It is therefore necessary to extract filaments by querying all Morse-Smale complexes within a given hierarchy.  } 
\label{fig:saddle_selection}
\end{figure}

Following this rationale, we translate this strategy into the use of queries that depend on the density properties of interest. 
Specifically, we query for filaments by specifying the 
density range $[M_b,M_e]$ of the clusters they connect (the maxima at the end points), 
as well as the density range $[S_b,S_e]$ of the lowest point along the connecting path (the density range of the 2-saddles). Figure~\ref{fig:saddle_selection} conceptually illustrates the characterization of filaments using density ranges, where density along filaments varies significantly necessitating simplification. 

Algorithm~\ref{alg:Select2Saddles} lists the algorithm to process such a query. The algorithm accepts, together with the combinatorial Morse-Smale complex $MSC$, the density ranges of 2-saddles $[S_b,S_e]$ and maxima $[M_b,M_e]$ as input (the subscripts $_b$ and $_e$ denote the beginning and ending of each density range respectively). 
The algorithm returns a list of 2-saddles that satisfy the above criteria together with the maximal Morse-Smale complex version in which they do so. 

\begin{varalgorithm}{\textsc{Select2Saddles}}
\caption{($[S_b,S_e]$, $[M_b,M_e]$ )}  
\begin{algorithmic}[1]
  \label{alg:Select2Saddles}
  \STATE $Sver$ = Empty Map
  \STATE $S = \{s | s $ is a 2-saddle,$ S_b < f(s) < S_e \}$
  \FORALL{ $s \in S $}
      \STATE $v_a = Max \{i | s $ is not simplified in $MSC_i \} $
      \STATE $v_b = Max \{i | t_i < M_b - f(s) \}  $      
      \STATE $v_c = Max \{i | s $ connects distinct Maxima in $ MSC_i \} $
      \STATE $Sver[s] = Min(v_a,v_b,v_c) $
  \ENDFOR  
  \STATE Sort S by Sver  
  \STATE $Ssel$ = Empty Set 
  \FORALL{ $s \in S $}
      \STATE $i = Sver[s]$
      \STATE $m_a,m_b = $ Maxima connected to $s$ in $MSC_i$
      \IF { $M_b < f(m_a),f(m_b) < M_e$}
	  \STATE insert $(s,Sver[s])$ in $Ssel$
      \ENDIF
  \ENDFOR  
  \RETURN $Ssel$
\end{algorithmic}
\end{varalgorithm}

The algorithm begins by creating a list $S$ of 2-saddles that have their function value in the given 2-saddle range $[S_b,S_e]$. Then, for each 2-saddle in $S$, a Morse-Smale complex version in which it possibly connects two maxima within $[M_b,M_e]$ is computed. The appropriate version is given by the minimum of three version indices $v_a,v_b$ and $v_c$. 

The version index $v_a$ is the finest Morse-Smale complex version in which the 2-saddle $s$ 
survives. In other words, $s$ is canceled in $MSC_{v_a+1}$ but not in $MSC_{v_a}$. 
This is pre-computed by examining the cancellation sequence. 
The version index $v_b$ corresponds to the last Morse-Smale complex version at which the 2-saddle $s$ connects two maxima, both with function value less than $M_e$.
This is possible because in successive versions of the Morse-Smale complex, the maxima connected to a 2-saddle via the same arc form an increasing sequence in terms of their function value. 
Thus, in the version where the absolute difference in function value of the last canceled pair is less than $M_b - f(s)$, 
the 2-saddle $s$ still possibly connects two maxima with function value less than $M_e$. 
The version index $v_c$ is the last Morse-Smale complex version at which the 2-saddle $s$ separates distinct maxima. In other words, it is not a strangulation in $MSC_{v_c}$. 
As a consequence of the cancellation preconditions, once a strangulation is created by a 2-saddle, it may be destroyed only by  canceling the 2-saddle with a 1-saddle. 
Thus there exists a maximal version index $v_c$ after which the 2-saddle remains connected to a single maximum. 
The version index $v_c$ is $-1$ when the 2-saddle is a strangulation in the initial Morse-Smale complex. 
In this case, the 2-saddle is not considered in further steps and is removed from $S$. 
Again, this is easily pre-computed for each 2-saddle by examining the cancellation sequence. 

The 2-saddles in the set $S$ are sorted based on their version indices.
This is done to optimize switching between the required Morse-Smale complex versions. 
Next, each 2-saddle $s$ in $S$ is checked to see if it separates two maxima within the maxima density range $[M_b,M_e]$. 
The list of 2-saddles that fulfill all of the above criteria is returned together with the associated version number of each 2-saddle.
This above list of 2-saddles is used to extract the filament geometry. Specifically, the ascending manifold of each 2-saddle is extracted from the corresponding version of the Morse-Smale complex. 
This may be done efficiently using the {\it cancellation merge DAG} data structure discussed by Gyulassy \etal{} \cite{Gyulassy12}. 

In some situations, it is desirable to perform some simplification to eliminate Poisson noise introduced due to meshing the domain. 
In these cases, a global simplification specifically for noise elimination, can be optionally introduced. Specifically, Algorithm~\ref{alg:Select2Saddles} returns only those 2-saddles that survive in hierarchical Morse-Smale complex versions above a specified threshold $T_s$, where $T_s$ is specified as a normalized fraction of the range of log-density values (normalized to $[0,1]$). $T_s$ is set to $0.0$ unless specifically mentioned. Similar to the inputs of Algorithm~\ref{alg:Select2Saddles}, $T_s$ may be updated during run-time.


\subsection{Volume Rendering}
\label{subsec:vol_enhancement}

We use volume visualizations of the density field to aid selection of parameters for Algorithm~\ref{alg:Select2Saddles}. 
The geometry of the selected filaments using Algorithm~\ref{alg:Select2Saddles} is superimposed upon a volume rendering of the density field. 
Based on the visualization of the extracted filaments and the density volume rendering, the parameters may be adjusted so that the structures correspond with the density volume rendering. 
Figure~\ref{fig:vk_new} shows an example of the overlay of the volume visualization with the selected structures. 
Direct volume rendering of the density is often not effective for visualization because of the formation of clusters at multiple scales. 
Furthermore, these clusters are often spatially far removed from the features of interest. 
In the supplemental document, we describe an application of our framework to suppress the opacity of regions spatially far removed from filamentary features of interest, which leads to a feature based volume rendering enhancement.

\section{Model Description}
\label{sec:modeldescription}
 
In this section, we briefly describe the models used to test our 
filament detection routine. These are the Voronoi evolution models 
and $\Lambda$CDM cosmological simulations. 

The Voronoi models provide us with vital quantitative information on the 
sensitivity of Felix to anisotropic filamentary patterns in the galaxy 
distribution. To this end it is of key importance that the Voronoi models have 
an a-priori known population fraction in different morphological elements: clusters, 
filaments, walls and voids. This makes them perfect test models for evaluating 
success and failure rates of the various identification methods. 

Although they involve filaments with a broad distribution of densities, the Voronoi 
models do not incorporate the multi-scale web-like patterns seen in 
realistic cosmological scenarios. To assess this aspect of the cosmic 
mass distribution, we turn to simulations of structure formation in 
the standard $\Lambda$CDM cosmology. Implicitly these include all relevant 
physical and dynamical processes of the evolving cosmic dark matter 
distribution. 
However, as we have no control over all aspects of the emerging mass 
distribution in $\Lambda$CDM simulations, for testing purposes they are not as 
informative as  Voronoi models~\footnote{In the literature, 
several studies use simplistic models using Voronoi tessellations. The models 
we use here are considerably more sophisticated, and represent a rather 
realistic depiction of the cosmic web in void-dominated cosmologies, 
see e.g. \cite{weykamp1993,shethwey2004,weygaert2007,aragon2013}.}

\subsection{Voronoi evolution models} 

\label{subsec:voronoi-models}

The Voronoi evolution models are a class of heuristic models for 
cellular distributions of galaxies that mimics the evolution of the 
Megaparsec universe towards a weblike pattern. They use Voronoi 
tessellations as a template for distribution of matter and related galaxy 
population \cite{weyicke1989,weygaert1991,weygaert2007}, and its 
subsequent evolution. 

In these models, one begins by fixing an underlying Voronoi skeleton, 
defined by a small set of randomly distributed nuclei in the 
simulation box. One then superposes a set of $N$ randomly distributed 
particles on this skeleton. The resulting spatial distribution of particles 
in the model is obtained by projecting the initially random distribution of 
particles on to the faces, edges, and nodes of the Voronoi 
tessellation. This results in a pattern in which one can distinguish four 
structural components: \emph{field} particles located in the interior 
of Voronoi cells, \emph{wall} particles within and around the Voronoi 
faces, \emph{filament} particles within and around the Voronoi edges 
and \emph{cluster} particles within and around the Voronoi nodes. 

One particular class of Voronoi clustering models are the Voronoi 
kinematic models, which seek to approximate the dynamical evolution of 
the large scale cosmic mass distribution. These models involve a 
continuous flow of galaxies towards the nearest wall, along a filament 
at the wall's edge, and subsequently towards the final destination, a 
vertex of the Voronoi tessellation. This motion is regulated by the increase 
of mean distance between the galaxies, an expression of void expansion 
and evacuation as a function of time.



The Voronoi models used in our experiments have $262,144$ particles 
distributed along the vertices, edges, faces and cells of the Voronoi 
skeleton in a box of side-length $200 \Mpch$. The skeleton is 
generated by $32$ randomly placed nuclei in the box. For the least evolved 
stage, most of the particles are in the cells, while for the most 
evolved stage, most of the particles are located in and around clusters. 
Table~\ref{tab:kinstat} presents the percentage distribution of particles 
in the various structural elements, as it changes with time. Stage 1 (dataset A) is 
the least evolved, while Stage 3 (dataset C) is the most evolved. The particles in 
and around the nodes, edges, and walls are Gaussian distributed around 
these elements, characterized by a thickness scale $R_f$ which is the 
standard deviation of the distribution. For our models, $R_f = 2 \Mpch$.
In the supplemental document, we present the volume renderings of the density distribution. 

\begin{table}
 \begin{center}
  \begin{tabular}{r|rrrr}
           &    cell  &  wall   & filament & cluster  \\ \hline
   A &   29.88\% & 43.57\% &  22.20\% &  4.33\%  \\
   B &    9.82\% & 32.13\% &  38.62\% & 19.42\% \\
   C &     3.5\% & 16.50\% &  28.70\% & 51.30\%  \\
  \end{tabular}
 \end{center}
 \caption{The relative abundance of particles in each structural element throughout the course of evolution. 
 }
 \label{tab:kinstat}
\vspace{-15pt} 
\end{table}

\subsection{$\Lambda$CDM cosmological simulations}

\label{subsec:lcdm-models}

The $\Lambda$CDM simulations are fully physical models that trace the distribution 
and evolution of dark matter in the universe based on current understanding of 
real physical laws. Dark matter is the gravitationally dominant matter component 
in the Universe and constitutes the major fraction of matter. As it 
is known to only interact gravitationally, modeling the behavior of 
dark matter is computationally fast and efficient. 
Such dark matter simulations form one of the principal tools towards 
understanding the evolution of the matter distribution in the Universe. 

The cosmological simulations that we used follow the standard $\Lambda$CDM cosmology. 
In this model, the matter content of this Universe is dominated by 
collisionless \emph{cold dark matter} (CDM) particles. The biggest contribution 
to the energy content of this Universe comes from dark energy, in 
the form of the cosmological constant $\Lambda$ (see \cite{WMAP5}),   
which drives its accelerated expansion at the current epoch. 


To present the results of our visual exploration framework, we use the Cosmogrid
simulations \cite{IRMP13}. It is a suite of simulations in a box of size $21 h^{-1} Mpc$, 
each differing in the number of particles. 
The particular simulation we use for our study comprises of $512^3$ particles. 
This is a relatively small scale in the context of the cosmic web. 
The mass resolution achieved is $8.21 \times 10^6 M\odot$. 
The initial conditions are setup at $z = 65$ using the Zel'dovich approximation 
\cite{Pancake1970}. 
The log-density field is available on a $128\times128\times128$ structured grid.  

Particularly characteristic in the evolving mass distribution of the 
Cosmogrid simulation is the large central under-density, surrounded 
by a range of smaller voids near its outer edge. In combination with 
its extremely high spatial resolution and state-of-the-art dynamic range, 
this renders the Cosmogrid simulation uniquely suited as a testbed for a 
case study of the internal structure of voids. It was precisely this 
circumstance that formed the rationale behind its exploitation in a 
previous study of the formation of dark halos along tenuous 
void filaments \cite{rieder2013}.

%
%


\section{Results and Discussion}
\label{sec:Experiments}
In this section, we demonstrate and discuss the salient features and 
potential applications of Felix. First, we evaluate the filaments extracted using Felix 
and compare with those extracted using MMF, SpineWeb, and {\rm DisPerSE} 
using the Voronoi kinematic datasets.  
Next, we present a visual exploration of the filaments in the {\rm Cosmogrid} simulation. 
The methods described above were implemented and tested on a computer 
system with an Intel Xeon(R) 2.0 GHz CPU and 8GB of RAM. 
In the supplemental document, we briefly discuss the structure finders against which we compare Felix.

\subsection{Filaments in the Voronoi model: a comparison}
Here we present an analysis of the filamentary structures extracted using 
Felix, and compare the results with the techniques detailed above. The comparison 
study concerns the analysis results obtained for the set of heuristic Voronoi 
evolution models described above. Since they are input parameters, in these models 
the classification of galaxies as void, wall, filament, and cluster are known 
a-priori. Following the application of one of the detection techniques we may then 
examine the validity and authenticity of the extracted structures by direct comparison 
with the true identity of a galaxy. 

For the comparison study, we define two measures. One quantifies the true detection 
rate of a method, the other the false identifications. We classify all galaxies within a 
distance $d$ from the extracted structures to be filament and cluster particles 
and the others to be void and wall galaxies. For a given set of structures extracted 
from a given dataset and a distance $d$, the true positive classification 
rate $TP_d$ is defined as
 
$$ TP_d = \frac{\text{\# filament and cluster galaxies correctly classified} }{ \text{\# filament and cluster galaxies }}.$$

Similarly, the false positive classification rate $FP_d$ is defined as  

$$ FP_d = \frac{\text{\# filament and cluster galaxies incorrectly 
classified} }{ \text{\# filament and cluster galaxies }}.$$ 

A large separation between these two measures indicates good discriminatory 
power of the classifier, and thus the proximity of relevant galaxies to 
the extracted structures. 

As we discuss in more detail below, the Felix's true and false detection 
rates are comparable, and in some situations better, than those obtained 
by {\rm DisPerSE}, SpineWeb and Nexus/MMF. 
A brief description of SpineWeb, and Nexus/MMF is provided in the supplemental document. As the {\rm DisPerSE} methodology is closely related to Felix, in the following paragraphs, we briefly describe it and contrast it against Felix.

\begin{figure*}
\centering
\subfloat[]{\label{fig:vk_new_2}\includegraphics[width=0.30\textwidth]{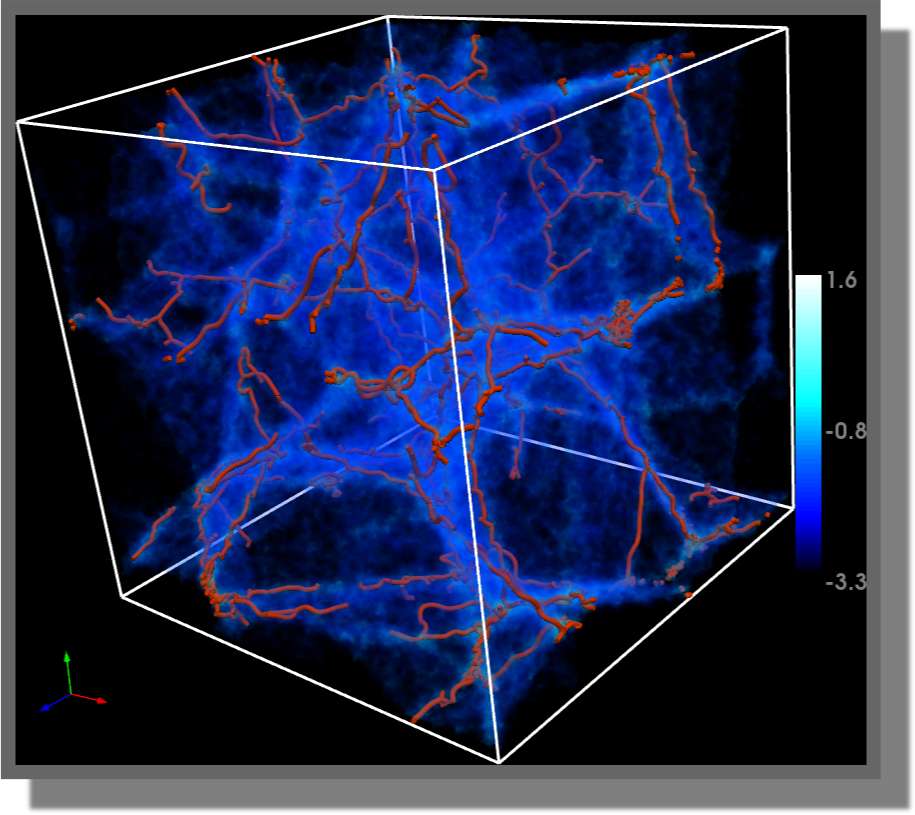}}
\hspace{0.02\textwidth}
\subfloat[]{\label{fig:vk_new_3}\includegraphics[width=0.30\textwidth]{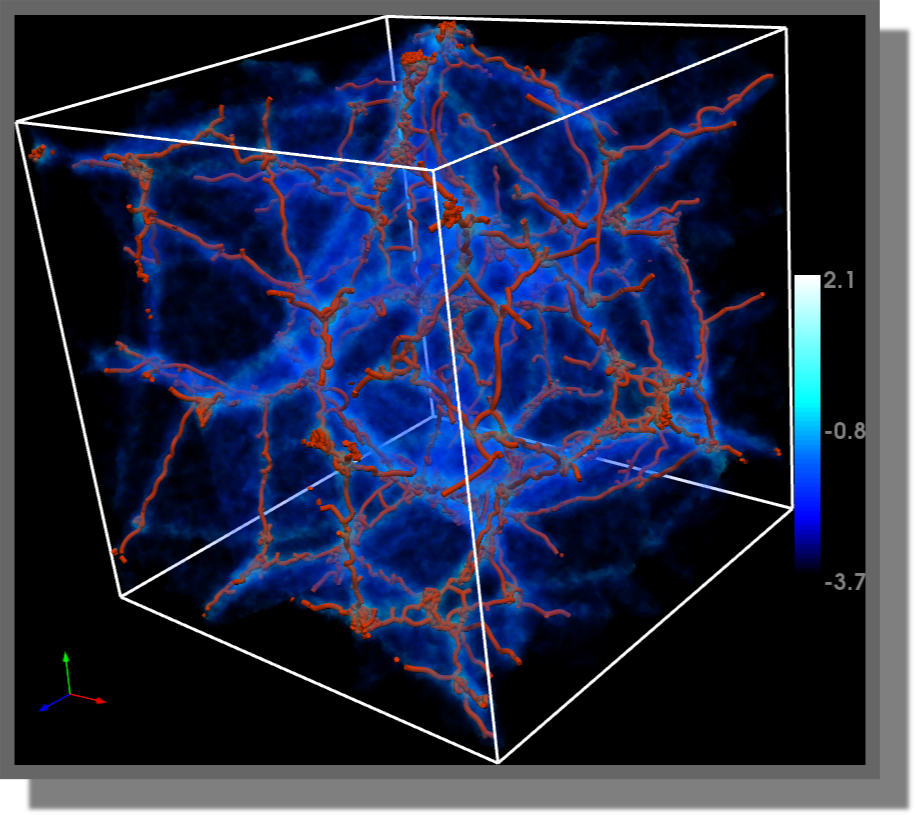}}
\hspace{0.02\textwidth}
\subfloat[]{\label{fig:vk_new_4}\includegraphics[width=0.30\textwidth]{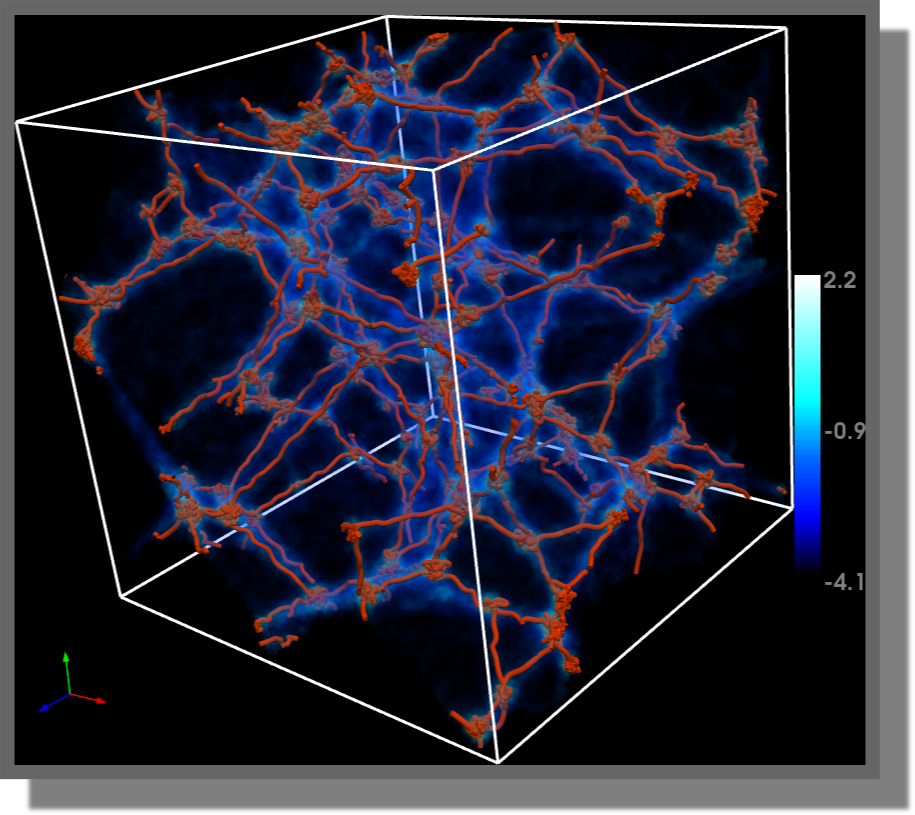}}
\caption{Filamentary structures extracted from datasets A, B, and C of the Voronoi evolution time-series using Felix with parameters for 
Algorithm~\ref{alg:Select2Saddles} set as follows: (a) 
$[S_b,S_e]=[10^{-1.6},\infty]$ and $[M_b,M_e]=[10^{-0.1},\infty]$, (b,c) $[S_b,S_e]=[10^{-1.6},\infty]$ and $[M_b,M_e]=[10^{0},\infty]$. 
Filaments are shown as orange tubes along with a volume rendering of the log-density field. 
The dense knot like structures show filaments within cluster-like regions. 
}
\label{fig:vk_new}
\end{figure*}

\begin{figure*}
\centering
\subfloat[]{\label{fig:vk_new_2_nsig5}\includegraphics[width=0.30\textwidth]{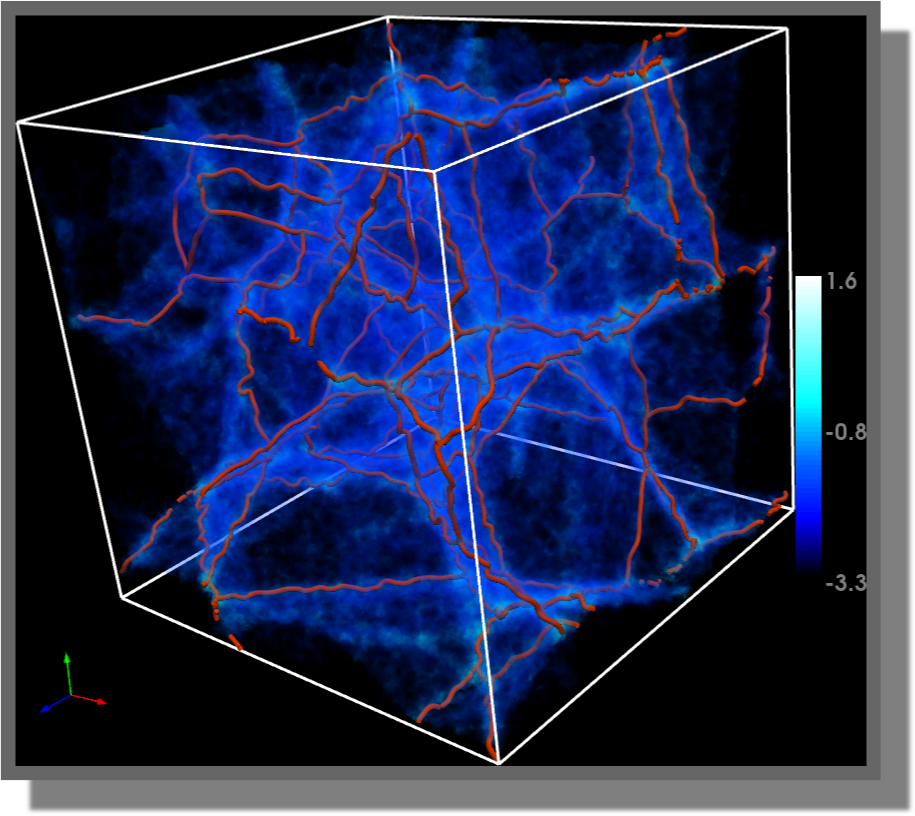}}
\hspace{0.02\textwidth}
\subfloat[]{\label{fig:vk_new_3_nsig5}\includegraphics[width=0.30\textwidth]{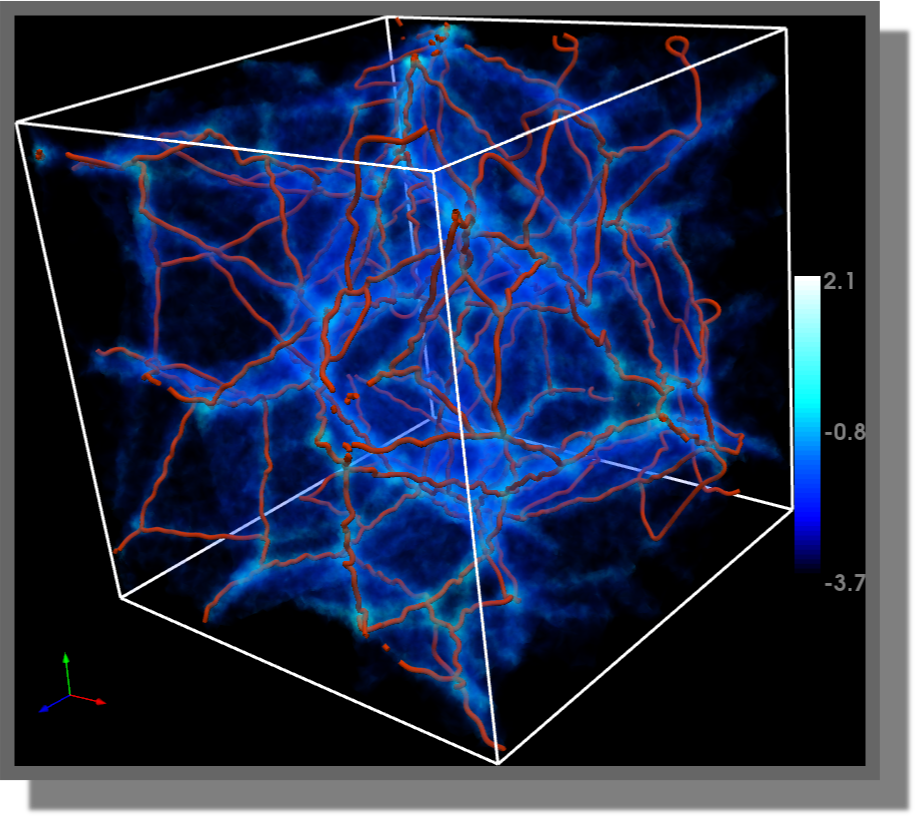}}
\hspace{0.02\textwidth}
\subfloat[]{\label{fig:vk_new_4_nsig5}\includegraphics[width=0.31\textwidth]{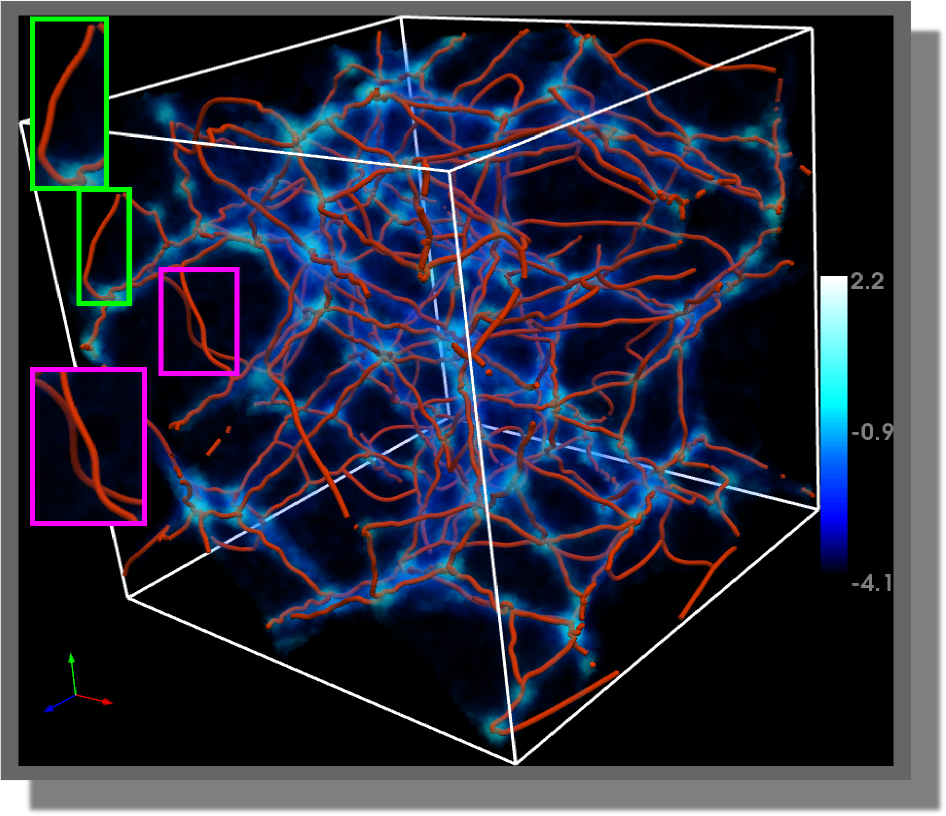}}
\caption{ 
Filamentary structures extracted from datasets A, B and C of the Voronoi evolution time-series using {\rm DisPerSE} with significance level of 
$5\sigma$. Filaments are shown as orange tubes along with a volume rendering of the log-density field.
The inset pictures show identified filaments that are within wall-like and void-like regions of the Voronoi kinematic datasets. 
}
\label{fig:vk_new_disperse}
\vspace{-15pt}
\end{figure*}

\paragraph*{\textbf{Felix}}
Figure~\ref{fig:vk_new} shows the extracted filaments for the Voronoi 
kinematic datasets $A$, $B$, and $C$ using Felix. The input density 
range parameters for Algorithm~\ref{alg:Select2Saddles} are selected 
interactively, using the visualization information of the procedure. 
Each update is accomplished within 2-3 seconds. This enables interactive 
visual feedback so that parameters may be adjusted further in subsequent 
iterative steps. 

\paragraph*{\textbf{Comparison A: Felix and DisPerSE}}

{\rm DisPerSE} \cite{Sousbie1,Sousbie2} is a closely related structure finder that also uses the Morse-Smale complex of the logarithm of the density field. It simplifies the  Morse-Smale complex using Topological Persistence. 
The function values are normalized by the rms of the density field fluctuation with respect to the mean, and the significance level for simplification is quoted in this unit. 
Felix is closely related to {\rm DisPerSE} as both use the Morse-Smale complex of the log-density field and involve feature extraction from it. 
{\rm DisPerSE} defines significant features as only those that remain unsimplified using the user defined significance threshold. 
It ignores the density range characteristics of the extracted features. 
A significant consequence is that filaments within void-like regions and cluster like regions are ignored/simplified away. If they are retained, then the mixing of features causes visual clutter. Furthermore, the significance parameter selection is a fixed constant and visual interaction plays no role in its selection. 
In contrast, given the ubiquity of filaments in various density regimes, Felix allows for density ranged based probes into filaments, within clusters and voids. Furthermore, the visual interactive aspect allows for user engagement in parameter selection, which is crucial for the set of features identified. Another difference is that Felix uses simplification only for noise removal and not feature identification.

In this experiment, we demonstrate the consequences of not correlating the density characteristics for filament extraction. 
Specifically, we demonstrate that the filaments extracted using Felix are more spatially proximal to filament and cluster particles in the Voronoi Kinematic simulation. 
Furthermore, we show that tuning the significance parameter is not a sufficient mechanism to extract the desired filaments in this dataset. 
In the next experiment, we demonstrate the exploration of filaments within high-density cluster like regions and low-density void like regions. 
Such a delineation, coupled with the visual exploration process, is not possible using {\rm DisPerSE}.


\begin{figure}
\centering
\includegraphics[width=0.4\textwidth]{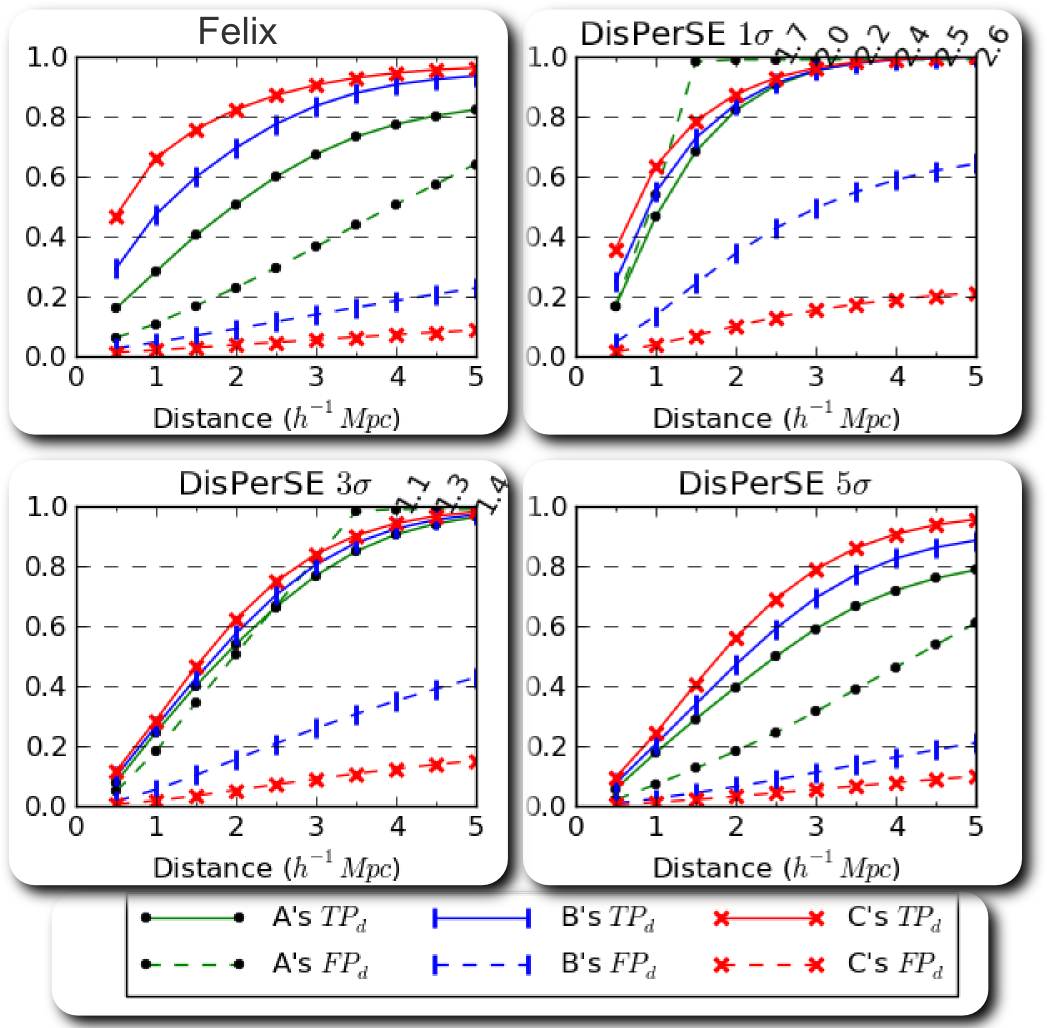}
\caption{Classification recovery rates for Voronoi kinematic datasets A, B, and C. 
Top left: Felix; top right: {\rm DisPerSE}, $1\sigma$ significance level; bottom left: 
{\rm DisPerSE}, $3\sigma$ significance level,
and (bottom right) {\rm DisPerSE} with a $5\sigma$ significance level.
False positive rates greater than 1.0 are clipped and respective values are shown.}
\label{fig:vk_new_tpfp_plot}
\vspace{-10pt}
\end{figure}

The recovery and failure rates for the Voronoi kinematic models $A$, $B$, and $C$ are shown 
in Figure~\ref{fig:vk_new_tpfp_plot}. The TP$_d$ and FP$_d$ of the Felix filaments are plotted 
in the top-left panel, the ones for {\rm DisPerSE} can be found in the top right-hand 
panel ($1\sigma$ significance threshold), bottom left-hand panel ($3\sigma$ significance 
threshold) and bottom-right panel ($5\sigma$ significance threshold). 

Felix shows good recovery rates for all datasets, particularly around $d=3\Mpch$. For the least 
evolved configuration $A$, and for locations where structures are least distinct, Felix 
still obtains moderately good recovery rates. The results for {\rm DisPerSE} 
with simplification thresholds $1\sigma$ and $3\sigma$ are comparable: at short 
distances the true detection rate is slightly lower than that of Felix, while at larger 
distances it performs marginally better. The situation is slightly different in the 
case of {\rm DisPerSE} with a $5\sigma$ simplification threshold. In the case of the 
more strongly evolved $B$ and $C$ datasets, {\rm DisPerSE} and Felix have 
similar false detection rates~$FP_{d}$, while the true detection rates~$TP_d$ of 
Felix are consistently higher.

For the lesser evolved datasets $A$ and $B$, the false detection rates $FP_{d}$ for 
{\rm DisPerSE} quickly increase to rather large values. For simplification thresholds of 
$1\sigma$ and $3\sigma$ it even surpasses values of unity. This may indicate that in certain 
circumstances an automatic detection of filaments from the Morse-Smale complex runs the risk 
of over-determining the population of filaments, even after considerable simplification. While the 
problem is not so acute in the most evolved stage $C$, where the morphologies are well separated, 
direct simplification strategies may not always succeed in properly classifying all filament, 
wall and void regions in the more moderately evolved stages $A$ and $B$. 

Figure~\ref{fig:vk_new_disperse} shows the filaments in the three Voronoi models detected 
by {\rm DisPerSE}, with a simplification threshold of $5\sigma$. 
In comparison with the structures in Figure~\ref{fig:vk_new}, the knot like structures 
present in clusters are absent. 
This leads to the cluster particles being far away from the filament end points, and thus the reduced $TP_d$ rates of DisPerSE. 
In contrast, Felix's ranged query allows us to retain only the filaments in cluster like regions and those that connect these cluster like regions, leading to better $TP_d$ rates.
Also, many filaments found by DisPerSE are within the wall-like and void-like regions of the Voronoi Kinematic simulation. Some examples are highlighted using insets in Figure~\ref{fig:vk_new_disperse}. 
Again these are filtered out by Felix's ranged query, which is not directly possible in DisPerSE. 
The inclusion of such structures in DisPerSE leads to its higher $FP_d$ rates. 

These findings suggest that a structure identification strategy based on a direct simplification 
procedure of the Morse-Smale complex should be applied with care to the density regimes being studied. 
Specifically, the superior classification rate profiles confirm that, using Felix, we can easily extract filaments that are spatially more proximal to the cluster and filament particles in the Voronoi Kinematic datasets. 
This issue is not easily addressed by the significance threshold of DisPerSE. In contrast, Felix provides an intutive density based handle to extract the desired features. 
Also, the possibility of having $FP_d$ value larger than $1$ in extreme situations is indicative of over-detection of filaments. 
This is potentially cumbersome for the analysis of genuine cosmological 
simulations and observational surveys. In more complex realistic circumstances, cosmic 
structure involves features over a wide range of densities and scales and structural 
morphologies that are not as well separated as in the simpler Voronoi models.

\begin{table}
 \begin{center}
  \begin{tabular}{|c|c|c|c|} \hline
  Dataset&              &    $TP_d$   &  $FP_d$  \\ \hline\hline

A &  ${Felix (d=3\Mpch)}$             &     0.67  &  0.36     \\ \hline
A &  ${DisPerSE (5\sigma,d=3\Mpch)}$ &     0.59  &  0.31     \\ \hline\hline 
B &   ${Felix (d=3\Mpch)}$             &     0.84  &  0.13     \\ \hline
B &   ${DisPerSE (5\sigma,d=3\Mpch)}$ &     0.69  &  0.11     \\ \hline
  &  ${Nexus/MMF}$                         &     0.85  &  0.13     \\ \hline\hline
C &   ${Felix (d=3\Mpch)}$             &     0.90  &  0.05     \\ \hline
C &   ${DisPerSE (5\sigma,d=3\Mpch)}$ &     0.78  &  0.05     \\ \hline
  & ${SpineWeb}$                    &     0.87  &  0.10     \\ \hline               
  \end{tabular}
 \end{center}
 \caption{Recovery rates of galaxies within $3 \Mpch$ of structures extracted using Felix compared with Nexus/MMF, 
SpineWeb, and {\rm DisPerSE} ($5\sigma$ significance level). 
}
 \label{tab:recovery_rates}
\vspace{-15pt}
\end{table}

\paragraph*{\textbf{Comparison B: Felix and SpineWeb}} 
For the comparison of the Felix and Spineweb \cite{Aragon2010}, we use a Voronoi Evolution model realization that 
is comparable to the advanced state of dataset $C$. We use the test result reported in \cite{Aragon2010} with 
respect to the model that has a similar percentage of particles in the four morphological features. In these model 
realizations, the clusters, filaments and walls have a Gaussian density profile with a scale of  
$R_g = 1\Mpch$. The spine has an effective width $d=2 R_g$ as the identified structures are thickened 
by $1$ voxel, with a size of $R_g = 1\Mpch$. 

For this configuration, Aragon-Calvo \etal{} \cite{Aragon2010} report true and false detection rates of $TP_d=0.87$ and $FP_d=0.10$ 
(see Table~\ref{tab:recovery_rates}). Felix attains the same recovery rate of $TP_d = 0.87$ at a smaller distance 
$d = 3\Mpch = 1.5 R_g$. For the same configuration, the failure rate parameter, $FP_d = 0.05$, 
is comparable to that reported for Spineweb. By comparison, at $d=2 R_g$, for Felix 
the recovery rates are $TP_d = 0.93$ and $FP_d = 0.07$. 

In summary, these results appear to suggest that Felix performs as well as SpineWeb.

\paragraph*{\textbf{Comparison C: Felix and Nexus/MMF}}
Since Nexus/MMF concerns a sophisticated formalism based on a 
scale space analysis, the parameters of detection do not have a direct 
correspondence with topology based techniques like {\rm DisPerSE}, {\rm SpineWeb}, and {\rm Felix}. 

We use dataset $B$ for a comparison with Nexus/MMF. This dataset is 
similar to the least evolved dataset used by \cite{aragon07b} in an 
evaluation of the MMF, the original density field based Nexus/MMF 
implementation. For similar values of the detection rate $TP_d$, 
both Felix and MMF have identical failure rates $FP_d$. This indicates 
that both procedures have a comparable detection behavior. 

\subsection{Filament Exploration}
\label{sec:fil_explore}

\begin{figure*}
\centering
\includegraphics[width=0.8\textwidth]{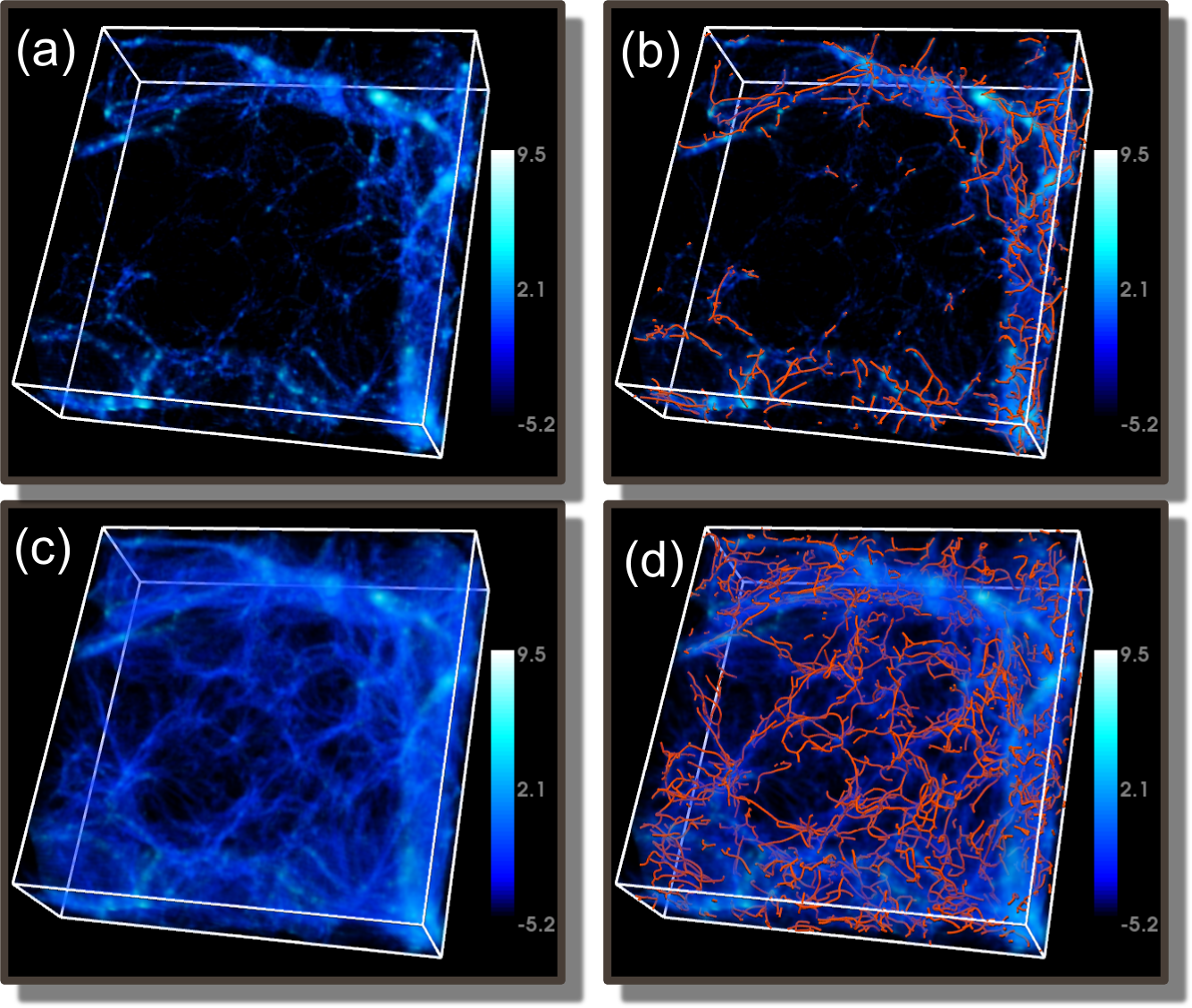}
\caption{Exploring filaments in high-density cluster-like environments and low-density void like regions in the Cosmogrid dataset. The selected region, comprising of a 3D region of z-slices from 69 through 105, contains filamentary structures in both types of environments. 
Volume rendering of the density field with opacity adjusted to highlight filaments in 
(a)~high density cluster-like regions, and (c)~low density void-like regions.
(b)~Filaments within high density cluster-like regions extracted with parameters 
$[S_b,S_e]=[10^{0},10^{9.6}]$ and $[M_b,M_e]=[10^{2.3},10^{9.6}]$. 
(d)~Filaments within low density void-like regions extracted with parameters $[S_b,S_e]=[10^{-2.5},10^{0.5}]$ and $[M_b,M_e]=[10^{0.5},10^{3.5}]$. For both sets of filaments, the value of $T_s$ is set to $0.05$.}
\label{fig:cosmo_grid}
\vspace{-15pt}
\end{figure*}

In this section we discuss the application of Felix to explore different classes of filaments from cosmological 
simulations.
The ability to filter filaments on a combination of morphological and density properties is helpful in situations where we wish to focus on, for example, the properties of galaxies residing in filaments in low-density void regions or in the high-density outskirts of clusters. 
This ability of Felix to identify a specified population of intravoid filaments or 
cluster inflow channels provides us with a microscopic instrument that allows a detailed and systematic exploration of 
the fine structure in the hierarchy of cosmic structure.
Felix is able to zoom in on such regions and delineate their detailed infrastructure. 

As the criteria for the identification of filaments and other web-like features still differ substantially between the 
various available techniques, the visual interaction aspect of Felix is a major practical asset 
in obtaining a proper user-defined selection of filaments. To this end, we may also point out that 
available automatic detection techniques may produce significant spurious results, which may substantially 
influence the results of targeted studies as the one illustrated here. A telling example of this has been 
discussed in the previous section. 

Figures~\ref{fig:cosmo_grid}a and \ref{fig:cosmo_grid}c present volume renderings of a 3D region of the Cosmogrid dataset ranging from z-coordinates $69$ to $105$. The bounding box of the dataset is $128\times128\times128$. 
This region is selected as it contains a large void like region surrounded by a large number of high density regions. 
The transfer function opacities have been adjusted to highlight the filament like structures in cluster-like and void-like regions respectively. 
The hierarchical Morse-Smale complex computation and the filament selection are executed on the entire dataset. The resulting filaments and volume rendering are clipped to the above mentioned region of interest.
The filament selection parameters of algorithm~\ref{alg:Select2Saddles} are adjusted interactively, via a  visualization step (see accompanying video). Each selection query takes approximately 1 second to process. The subsequent 
extraction of filament geometry depends on the number of selected 2-saddles. This takes approximately 
4 seconds. Thus, the query framework may be used to interactively change parameters and visually correlate the set of extracted features with the underlying density distribution. 

The filament selections obtained following the application of the interactive procedure are 
illustrated Figures~\ref{fig:cosmo_grid}b and \ref{fig:cosmo_grid}d. 
Each shows filaments in a different environmental density regime. 
Figure~\ref{fig:cosmo_grid}b shows filaments that exist near and within the high density cluster 
like regions.
These are the filaments that form the spine of the cosmic web. 
Figure~\ref{fig:cosmo_grid}d shows filaments within void-like regions. 
The combination of density criteria and interactive visualization enables us to zoom 
in on this system of intra-void filaments. 
They are the faint residuals of the smaller-scale filaments that constituted the spine 
of the cosmic web at earlier cosmic epochs, and as such represent a direct manifestation of the hierarchical 
buildup of cosmic structure. At the current epoch, the intra-void filaments appear to define a different pattern than the prominent 
filamentary bridges between clusters of galaxies. As a result of the tidal influence of surrounding large-scale mass concentrations 
they are conspicuously aligned along a direction correlated with the main axis of the embedding void.



\section{Conclusion}
\label{sec:conclusion}
We have presented a Topology based Framework, named Felix, to probe filament structure in the large scale universe. 
The framework is particularly designed to probe filamentary structures in different density regimes, and optimally preserve structural detail in regimes of interest.
While other cosmic structure analysis tools do not include a facility to select web-like
features according to tailor-made aspects and characteristics, this is precisely the mission of
the Felix procedure. 
We directed Felix towards a case study of the  the filamentary infrastructure
and architecture of cosmic voids and demonstrated that it successfully extracts the network
of tenuous filaments pervading their interior \cite{weykamp1993,shethwey2004,weygaert2007,aragon2013} .


In an accompanying study, we plan to exploit the Felix facility to study the physical characteristics of the extracted samples of intra-void filaments. This also involves their halo and subhalo population, their gas content, 
and the relation of these with the embedding voids and surrounding large-scale mass distribution. This will be 
of key importance towards understanding the formation and evolution of void galaxies \cite{wey11,kreckel11,kreckel12} 
and specifically that of the issue of the missing dwarf galaxies \cite{peebles2001}. In addition, following the 
recognition that void architecture represents a potentially sensitive probe of dark energy and dark matter and a keen 
test of modified gravity theories \cite{ryden96,ryden01,PL07,LW2010,BWDP12,LW12,clampitt13,SPWW14}, the filament samples 
extracted by Felix will be subjected to a systematic study of their dependence on cosmological parameters.

As an immediate extension, we plan to use Felix with other scalar fields such as the tidal force field. Another possible direction is the visualization and analysis of the hierarchy of voids, walls, and filaments in cosmological datasets. 
Interactive visual exploration of these intricate structural networks remains a challenging and largely unexplored problem of major significance.

\section*{Acknowledgment}
Vijay Natarajan would like to acknowledge funding from Department of Science and Technology, India. Grant No. SR/S3/EECE/0086/2012.
Rien van de Weygaert acknowledges support by the John Templeton Foundation
grant nr. FP5136-O.


\ifCLASSOPTIONcaptionsoff
  \newpage
\fi



\bibliographystyle{IEEEtran}
\bibliography{VisTopoGeomOfCosmicWebSims}

\vspace{-40pt}
\begin{IEEEbiography}[{\includegraphics[width=1in,height=1.2in,clip,keepaspectratio]{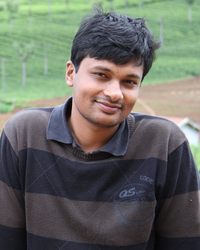}}]
{Nithin Shivashankar} is a doctoral candidate at the Department of Computer Science and Automation, Indian Institute of Science, Bangalore. He holds a bachelors degree in computer science from the Manipal Institute of Technology. His research interests include scientific visualization and computational topology.
\end{IEEEbiography}

\vspace{-45pt}
\begin{IEEEbiography}[{\includegraphics[width=1in,height=1in,clip,keepaspectratio]{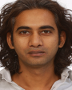}}]
{Pratyush~Pranav} is working for his PhD degree at the Kapteyn Astronomical Institute, Groningen, the Netherlands. He received a bachelors degree from St. Stephens college, Delhi, India, and holds a  Masters degree from the Department of Physics at the Indian Institute of Science, Bangalore, India. His research interests include computational topology, cosmology, large scale structures in the Universe and Scientific visualization. 
\end{IEEEbiography}

\vspace{-45pt}
\begin{IEEEbiography}[{\includegraphics[width=1in,height=1.2in,clip,keepaspectratio]{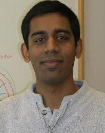}}]
{Vijay Natarajan} is an associate professor at the Indian Institute of Science, Bangalore. He received the Ph.D. degree in computer science from Duke University in 2004 and holds the B.E. degree in computer science and M.Sc. degree in mathematics from Birla Institute of Technology and Science, Pilani, India. His research interests include scientific visualization, computational geometry, computational topology, and meshing.
\end{IEEEbiography}

\vspace{-45pt}
\begin{IEEEbiography} [{\includegraphics[width=1in,height=1.2in,clip,keepaspectratio]{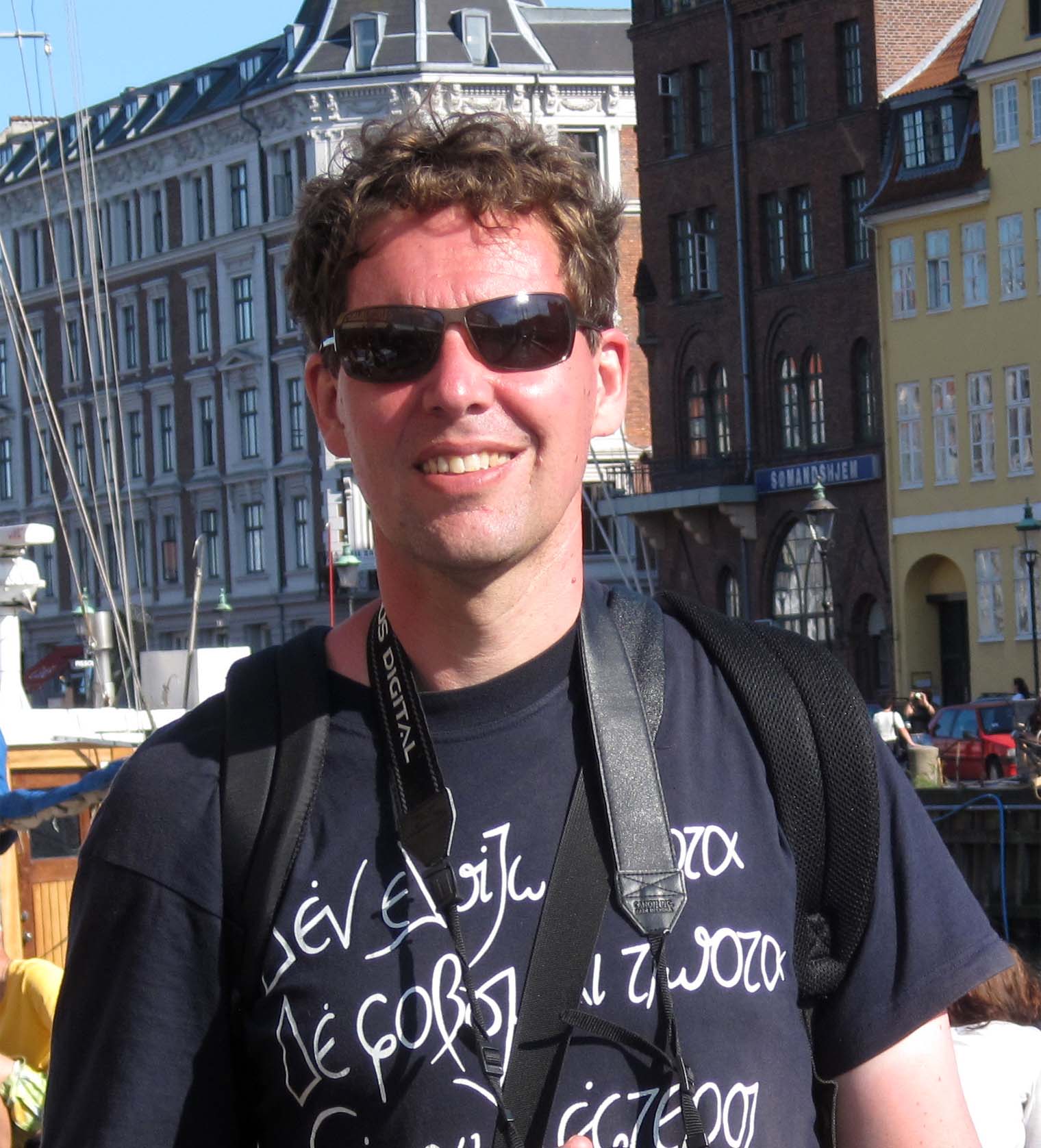}}]
{Rien~van~de~Weygaert} is professor of cosmic structure formation at the Kapteyn Astronomical Institute of the University of Groningen. He received his Ph.D. in astrophysics at Leiden University in 1991 on the role of cosmic voids in the large scale mass distribution in the Universe.
His research interests are cosmology, the large scale structure of the Universe, cosmic structure
formation, galaxy formation and evolution, computational geometry and topology, and the history of astronomy. 
\end{IEEEbiography}

\vspace{-35pt}
\begin{IEEEbiography}[{\includegraphics[width=1in,height=1.2in,clip,keepaspectratio]{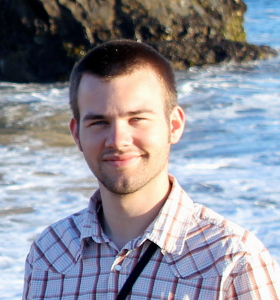}}]
{Patrick~Bos} is a PhD student at the Kapteyn Astronomical Institute of the University of Groningen in The Netherlands. He studies alternative models of dark energy through simulations and tries to find new probes to disentangle the nature of dark energy. He is also involved in building high-dimensional statistical models and techniques for inferring the initial conditions of the universe. 
\end{IEEEbiography}

\vspace{-45pt}
\begin{IEEEbiography}[{\includegraphics[width=1in,height=1.2in,clip,keepaspectratio]{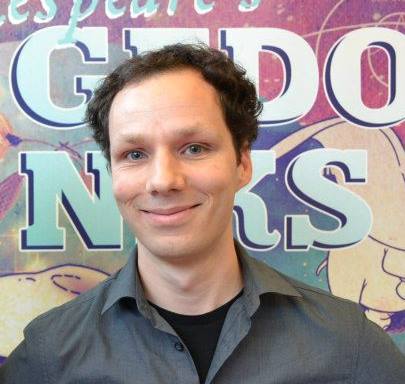}}]
{Steven~Rieder} is a postdoctoral researcher at RIKEN Advanced Institute for Computational Science, Japan. He received his Ph.D. degree in astrophysics at Leiden University in 2013, on the topic of multi-scale N-body simulations. His research interests include computational astrophysics, the large scale structure of the Universe, galaxy formation and evolution and the formation and disruption of star clusters.
\end{IEEEbiography}




\end{document}